\documentclass[pra,twocolumn,showpacs,letterpaper,superscriptaddress]{revtex4-1}
 \usepackage{graphicx,amsmath,amssymb,amsfonts,latexsym,xcolor,dcolumn,bm,epsfig,subfigure}
 \usepackage[plainpages=false,hyperfootnotes=false,colorlinks=false]{hyperref}
 \usepackage[normalem]{ulem}
 \usepackage{dsfont}
\usepackage{orcidlink}

 \allowdisplaybreaks[3]

\begin{document}

 \title{Quantum Thermodynamic Uncertainty Relations, Generalized Current Fluctuations and Nonequilibrium Fluctuation-Dissipation Inequalities}

 \author{Daniel Reiche\orcidlink{0000-0002-6788-9794}}
 \email{Corresponding author: reiche@physik.hu-berlin.de}
 \affiliation{Institut f\"ur Physik, Humboldt-Universit\"at zu Berlin, Newtonstr. 15, 12489 Berlin, Germany}

 \author{Jen-Tsung Hsiang\orcidlink{0000-0002-9801-208X}}
 \email{cosmology@gmail.com}
 \affiliation{Center for High Energy and High Field Physics, National Central University, Taoyuan 320317, Taiwan, ROC}

 \author{Bei-Lok Hu\orcidlink{0000-0003-2489-9914}}
 \email{blhu@umd.edu}
 \affiliation{Maryland Center for Fundamental Physics and Joint Quantum Institute, University of Maryland, College Park, Maryland 20742-4111 U.S.A.}

 \begin{abstract}
 Thermodynamic uncertainty relations (TURs) represent one of the few broad-based and fundamental relations in our toolbox for tackling the thermodynamics of nonequilibrium systems.
 One form of TUR quantifies the minimal energetic cost of achieving a certain precision in determining a nonequilibrium current.
 In this initial stage of our research program, our goal is to provide the quantum theoretical basis of TURs using microphysics models of linear open quantum systems where it is possible to obtain exact solutions.
In paper [Dong \textit{et al.}, Entropy {\bf 24}, 870 (2022)], we show how TURs are rooted in the quantum uncertainty principles and the fluctuation-dissipation inequalities (FDI) under fully nonequilibrium conditions.
In this paper, we shift our attention from the quantum basis to the thermal manifests.
Using a microscopic model for the bath's spectral density in quantum Brownian motion studies, we formulate a ``thermal'' FDI in the quantum nonequilibrium dynamics which is valid at high temperatures.
This brings the quantum TURs we derive here to the classical domain and can thus be compared with some popular forms of TURs.
In the thermal-energy-dominated regimes, our FDIs provide better estimates on the uncertainty of thermodynamic quantities. Our treatment includes full back-action from the environment onto the system.
As a concrete example of the generalized current, we examine the energy flux or power entering the Brownian particle and find an exact expression of the corresponding current-current correlations.
In so doing, we show that  the statistical properties of the bath and the causality of the system+bath interaction both enter into the TURs obeyed by the thermodynamic quantities.
 \end{abstract}

\maketitle

 \section{Introduction}\label{Sec:Introduction}

 Perhaps one of the most interesting recent developments in the pursuit to understand nonequilibrium sciences are the family of thermodynamic uncertainty relations \cite{barato15,gingrich16,dechant18a,barato19,saryal19,hasegawa20,hasegawa21,menczel21} (see also Refs. \cite{horowitz20,falasco20a,falasco22}).
 In one form, a TUR relates the fluctuations of a nonequilibrium current with the minimum of dissipation energy during the nonequilibrium process.
 Originally derived for classical, Markovian systems, the TUR has been quickly extended to over- \cite{gingrich17} and underdamped \cite{vu19} dynamics of a Brownian particle as well as Markovian quantum systems by means of large deviation methods \cite{carollo19} and the Cram\'er--Rao bound for the quantum Fisher information \cite{hasegawa20}.
 The latter technique has also been used to derive a connection between the TUR and fluctuation theorems for classical systems \cite{hasegawa19} and has just recently been used to derive an extension of the TUR for quite general open quantum systems \cite{hasegawa21}.
 We further mention that a TUR has been investigated for steady-state heat transfer \cite{saryal19,saryal21} and a similar expression could be found for the relation between the entropy production and the time it takes to complete a nonequilibrium process \cite{falasco20}.

 Aiming at finding the quantum roots of the uncertainties of thermodynamic quantities in nonequilibrium systems, we demonstrated \cite{dong22} that for Gaussian open quantum systems, thermodynamic functions are functionals
 of the Robertson-Schr\"odinger uncertainty (RSU) function.
 Using recent results on the nonequilibrium free energy and nonequilibrium effective temperature \cite{hsiang21}, we showed that a fluctuation--dissipation inequality (FDI) exists at all times in the nonequilibrium dynamics of the open system.
 In this sequel paper, we continue these veins of investigation and show how a thermodynamic uncertainty relation (TUR) for macroscopic quantities can formally be derived.
 While these are often motivated by phenomenological considerations in the literature, we show how some such relations can be obtained in rigorous ways within a microscopic quantum, even quantum field theory, framework.
 The centerpiece is the generalized current fluctuations.
 In short, from the mathematical perspective, while our first paper \cite{dong22} is concerned with physical quantities that are derived from second-order correlations functions, we now direct our attention to higher-order correlation functions and their physical imprints on Gaussian open quantum systems.
 Let us examine the ingredients one by one, and then look at their synthesis which leads to the aforementioned inequalities and uncertainty relations.

 \subsection{Fluctuations on Center Stage}

 Fluctuations fundamentally limit the precision of measurements.
 Fluctuation-induced effects, however, present themselves as a useful tool in controlling physical systems on the micro- and nanoscale.
 A precise understanding and manipulating of the fluctuations is hence crucial in the design of future technologies, especially in the strive for further miniaturization of devices (see, e.g., the recent reviews \cite{bongs19,frye21,gong21,reiche22} and references therein).

 Some fluctuations are introduced by operational techniques or imperfections and can in principle be eradicated by optimizing the experimental protocol.
 Other fluctuations are more fundamental and can never be circumvented.
 Possibly the most famous of such fluctuations are the quantum fluctuations whose physical implications are summarized by the quantum uncertainty principles (QUP) \cite{heisenberg27} and, in particular, the Robertson--Schr\"odinger uncertainty principle \cite{robertson29,schroedinger30}.
 In many realistic scenarios, the system of interest is also coupled to an environment.
 Thermal noise enters from a finite temperature bath, and a
 finite coupling strength between the open system and its environment can also be represented by noises.
 The nonequilibrium dynamics of the open quantum system can further modify the uncertainty relations \cite{hu93a,hu95,koks97}.

 In equilibrium, the corresponding variance of a quantum observable in an open system is accounted for by the fluctuation--dissipation theorem \cite{callen51,kubo66}.
 It states an exact relation between variance and dissipation due to a detailed balance between the average incoming and outgoing power \cite{li93}.

 \subsubsection{Nonequilibrium Relations}

Departing from equilibrium enriches the fluctuation spectrum with respect to the corresponding equilibrium situation.
 The search for the statistical description of both classical and quantum systems in nonequilibrium situations has instigated the development of a number of theorems.
 To name a few examples, we find the fluctuation theorems \cite{jarzynski96,crooks99,hatano01} (see, e.g., also the reviews of refs. \cite{esposito09,campisi11,seifert12,deffner19a}), the nonequilibrium fluctuation--dissipation relations for steady states \cite{eckhardt84,seifert10,intravaia14,lopez18,hsiang20a} (see also refs. \cite{maes20,maes20a,caprini21,caprini21a} for nonequilibrium fluctuation--dissipation relations of classical active systems), and the fluctuation--dissipation inequality \cite{fleming13,reiche20a}.
 The latter is a reflection of the nonequilibrium condition the system exists in for the full duration \emph{before} it reaches equilibrium.
 In simple terms, it states that the energy stored in the fluctuations of the environment (statistical operator) is always equal to or exceeds the energy dissipated into the environment \cite{fleming13,reiche21}.
 It can hence be understood as a generalization of fluctuation--dissipation relations which can often be shown to hold at late times \emph{after} a system has settled down to equilibrium while interacting with its environment.

 \subsubsection{Fluctuations of Generalized Current}

Due to their fundamental nature, it can be expected that the (often microscopically formulated) uncertainty principles are to some extent inherited by certain macroscopic observables.
 This line of thought draws attention to another perspective on uncertainties in nonequilibrium systems that has seen a surge of interest in the recent years, namely the \emph{thermodynamic uncertainty relation} (TUR) \cite{horowitz20}.
 Despite versatile contexts and examples (see beginning of current section), the particular relations can often be summarized in terms of a \textit{{generalized current}} $\hat{J}$. May it be, for example, the position of a particle \cite{barato15}, certain measurements on atomic systems \cite{hasegawa20}, or the exchanged energy in heat transfer \cite{saryal19}, the thermodynamic uncertainty relation states that the fluctuations of the generalized current $\langle\hat{J}^2\rangle$ are always larger than or equal to the average current $\langle\hat{J}\rangle$, with a thermal proportionality factor depending on the temperature of the heat bath(s). If the average current can be connected to dissipation from the system of interest into the environment, the TUR provides a nonequilibrium measure for the thermodynamic cost of achieving a desired precision on measurements of the current \cite{barato15}.

 \subsection{This Work---Key Findings and Organization}

 In the present manuscript, we aim to connect the microscopic picture of interacting  quantum systems (quantum uncertainty principles) with  the macroscopic picture of nonequilibrium thermodynamics (thermodynamic uncertainty relation) from a conceptual point of view.  To this end, based on the generic model for quantum Brownian motion, we first study the connection between fluctuations in the system and the bath spectral density. In this way, we extend previous work on the nonequilibrium fluctuation--dissipation inequality \cite{dong22} to something we call ``thermal fluctuation--dissipation inequality'' which additionally takes the impact of the bath temperature into account. Secondly, we calculate the fluctuations of the current of energy entering the system.
 We then combine the exact result for the current fluctuations with the thermal fluctuation--dissipation inequality to derive a combined inequality.
 This inequality incidentally turns out to formally represent a thermodynamic uncertainty relation, i.e., it is proportional to the average current times a thermal factor. In contrast to parts of the related literature, our result - showing the emergence of a thermodynamic uncertainty relation from the nonequilibrium evolution -fully incorporates the non-Markovian features of the system+bath interaction (exact in all orders of coupling).
 Since our formalism can be easily extended to the problem of heat transfer, we confirm the soundness of our results by reproducing a known TUR for steady-state heat transfer \cite{pagel13,saryal19}.
 It  turns  out  that  a  limited  number  of  assumptions  is needed  to  connect  the  two  worlds - quantum uncertainty principles and the TUR - which,  in  turn,  establishes  a  clear  hierarchy  of  different  inequalities  that are  connected  to  their  respective  physical  perspectives: from  microscopic  quantum  mechanics,  over  thermodynamic quantities, and all the way to concrete physical realizations.


This paper is structured as follows.
 We work in an open quantum system \mbox{framework~\cite{rammer07,breuer07}} and introduce the quantum stochastic dynamics of the system using the time-honored quantum Langevin equation of generic quantum Brownian motion in Section~\ref{Sec:StochDyn} \cite{schwinger61,feynman63,caldeira83,grabert88,hu92,hu93b,halliwell96,calzetta01,calzetta03,haenggi05}.
 In Section \ref{Sec:FDIRS}, we deduce the fluctuation-dissipation inequality and derive the space-momentum uncertainty.
 We relate the two to the power flow between system and environment as well as the total entropy production and the entropic uncertainty relation during the equilibration process in Section \ref{Sec:EnergyFlow}.
 We then specify the spectral density and the dissipation kernel characterizing the environment of the particle (see Section \ref{Sec:ThermalFDI}) and derive a thermodynamic uncertainty relation (see Section \ref{Sec:TUR}).
 Specifying the fluctuation--dissipation inequality to finite temperatures to include thermal noise contributions (see Section \ref{Sec:ThermalFDI}), we can highlight when thermodynamics enters the stage by building a concrete connection to the functional form of the thermodynamic uncertainty relation in Section \ref{Sec:TURHeat}.
 In Section \ref{Sec:HeatTrans}, we briefly show how our formalism can be applied to heat transfer and comment on the consistency with the respective (steady-state) TURs that have been derived earlier.
 We conclude our manuscript with a discussion in Section \ref{Sec:Summary}.


 \section{Stochastic Dynamics of Gaussian Systems: 2nd and Higher-Order Correlations}\label{Sec:StochasticDynamics}

 We begin with a brief review of stochastic dynamics of open quantum systems using the ubiquitous quantum Brownian motion model. We show from the Langevin equation how the open system's dissipative dynamics is linked to the fluctuations in its environment, registered in the dissipation (response) and noise (correlation) kernels. From this, we show how to obtain the second and fourth-order correlations of currents. This prepares us to tackle the main tasks we set forth in our goals.

  \subsection{Fluctuations and Stochastic Dynamics of Open Quantum Systems}\label{Sec:StochDyn}

 In order to introduce our system, we loosely follow refs. \cite{ford88,ford91,ford01,intravaia03,calzetta03,fleming13} and refer readers to the monograph \cite{calzetta08} for a more comprehensive discussions.

 We consider the dynamics of a single bosonic quantum degree of freedom $\hat{q}$ under the influence of a quadratic potential $V(\hat{q})=\omega_0^2\hat{q}^2/2$ with bare frequency $\omega_0$ \cite{Note1}.
 The system is coupled to a bosonic bath at finite temperature $T$ described by the generic environmental operator $\hat{E}$.
 We assume a simple bilinear coupling \cite{intravaia03}
 \begin{align}
 \hat{H}_I=-\hat{q}\hat{E},
 \end{align}
where we absorbed the coupling constant in the definition of $\hat{E}$.
 We note that this choice of coupling implicitly demands that the Hamiltonian is appropriately renormalized in order to account for the (Lamb-)shift of the potential's minimum frequency $\omega_0$ due to the interaction with the environment \cite{ford88,intravaia03} [see also Equation \eqref{Eq:QLE}].
 Under the assumption of a linear (Gaussian) environment, the bath operator can be expressed self-consistently as \cite{ford88,calzetta03,fleming13}.
 \begin{align}
 \label{Eq:BathOp}
 \hat{E}(t)
 	&=
 	\hat{\xi}(t)
 	-
 	2\int_{t_i=0}^t\mathrm{d}
 	\tau~
 	\Gamma(t,\tau)
 	\hat{q}(\tau),
 \end{align}
 where $\hat{\xi}$ is the unperturbed stochastic operator of the environment in absence of the system, and we set the initial time of the experiment to $t_i=0$.
 The integral in Equation~\eqref{Eq:BathOp} connects the impact of the system on the environment which, in turn, back-acts on the particle by means of the real-valued and causal response kernel $\Gamma(t,\tau)$.
 The noise spectrum is intimately connected to the response kernel.
 \begin{align}
 \label{Eq:NoiseSpecGen}
 \langle
 \hat{\xi}(t)\hat{\xi}(t')
 \rangle
 	&
 	=
 	\nu(t,t')
 	+
 	i\hbar
 	\Gamma(t,t')
 	,
 \end{align}
 where we average over the density matrices of the system $\hat{\rho}_S$ and the bath $\hat{\rho}_E$, which we assume to factorize at initial time, i.e., $\hat{\rho}(0)=\hat{\rho}_S(0)\otimes\hat{\rho}_E(0)$.
 The real-valued correlation function $\nu(t,t')$ and the kernel $\Gamma(t,t')$ are given by \cite{intravaia03}.
 \begin{subequations}
 \label{Eq:Comm}
 \begin{align}
 \nu(t,t')
 	&=
 	\frac{1}{2}
 	\langle\{\hat{\xi}(t),\hat{\xi}(t')\}\rangle
 	\equiv
 	\langle
 	\hat{\xi}(t)\hat{\xi}(t')
 	\rangle_{\text{s}},
 	\\
 \Gamma(t,t')
 	&=
 	\frac{1}{2i\hbar}
 	\langle
 	[\hat{\xi}(t),\hat{\xi}(t')]
 	\rangle,
 \end{align}
 \end{subequations}
 where $[\cdot,\cdot]$ is the commutator and $\{\cdot,\cdot\}$ the anticommutator of two operators, and we used that $\langle\hat{\xi}(t)\hat{\xi}(t')\rangle=\langle\hat{\xi}(t')\hat{\xi}(t)\rangle^*$.
 It follows a stochastic integro-differential quantum Langevin equation \cite{ford91,ford01}.
 \begin{align}
 \label{Eq:QLE0}
 \ddot{\hat{q}}(t)
 	+
 	2
 	\int_0^{t}\mathrm{d}\tau~
 	\Gamma(t,\tau)
 	\hat{q}(\tau)
 	+
 	\omega_0^2
 	\hat{q}(t)
 	=
 	\hat{\xi}(t)
 \end{align}
 which, for a given set of initial operators $\{\hat{q}(0),\dot{\hat{q}}(0)\}$, fully determines the dynamics of the system.
 For simplicity, we from now on consider stationary functions $\nu(t,t')=\nu(\tau=t-t')$ and $\Gamma(t,t')=\Gamma(\tau)$.
 Equation \eqref{Eq:QLE0} allows for an intuitive interpretation of the separate terms:
 due to the resemblance between Equation \eqref{Eq:QLE0} and the equation of motion for a randomly moving particle in a thermal viscous environment, it is usually referred to as \emph{{quantum Brownian motion}
 } \cite{caldeira83,grabert88,hu92}.
 $\hat{\xi}$ acts as noise  ``driving'' the system.
 The kernel $\Gamma$, on the other hand, encodes the dissipative processes, i.e., energy losses of the system to the environment, as well as the rescaling of the {bare} frequency $\omega_0$.
 The latter can be made explicit by defining the dissipation kernel.
 \begin{align}\label{Eq:DissipationDef}
 -\partial_\tau\gamma(\tau)=\Gamma(\tau)
 \end{align}
 which yields after a partial integration
 \begin{align}
 \label{Eq:QLE}
 \ddot{\hat{q}}(t)
 	+
   2
 	\int_0^{t}\mathrm{d}\tau~
 	\gamma(t-\tau)
 	\dot{\hat{q}}(\tau)
 	+
 	\tilde{\omega}_0^2
 	\hat{q}(t)
 	=
 	-\gamma(t)\hat{q}(0)
 	+
 	\hat{\xi}(t),
 \end{align}
 where we have redefined the resonance energy as
 $
 \tilde{\omega}_0^2=\omega_0^2-\gamma(0)
 $.
 In order to keep notation simple, we do not print the extra tilde in the following.
 Furthermore, in the remainder of the manuscript, we assume that $\langle\hat{q}(0)\rangle=\langle\dot{\hat{q}}(0)\rangle=0$.
 We note that, at the price of restricting to linear systems, the previous equation of motion is non-Markovian and general to all orders of coupling between system and environment, i.e., we are not limited by the Born--Markov or rotating wave approximation \cite{ford96a,ford97,fleming10}.
 It is well-known that the quantum Langevin equation [Equation \eqref{Eq:QLE}] can be solved for the canonical pair
 $
 \mathbf{Q}(t):=
 [\hat{q}(t)\;\dot{\hat{q}}(t)]^{\mathrm{T}}
 $  by means of the response function (see Appendix \ref{SolvLangevin} for details).
 \begin{align}
 \label{Eq:GreenFourier}
 \chi(t)
 	=
 	\int\frac{\mathrm{d}\omega}{2\pi}~
 	\alpha(\omega)e^{-i\omega t}
 \end{align}
 with
 $
 \alpha(\omega)=[\omega_0^2-\omega^2-2i\omega\gamma_{\theta}(\omega)]^{-1}
 $
 as the linear susceptibility and $\gamma_{\theta}(t)=\theta(t)\gamma(t)$ and $\theta(t)$ as the Heaviside step function.
 We note that we distinguish between a function and its Fourier transform only by the different argument.
 The corresponding second-order fluctuations are given by
 \begin{subequations}
 \label{Eq:FluctSystem}
 \begin{align}
 \langle\Delta\mathbf{Q}^2(t)\rangle_{\mathrm{s}}
 	&\equiv
 	\langle\mathbf{Q}^2(t)\rangle_{\mathrm{s}}
 	-
 	\langle\mathbf{Q}(t)\rangle_{\mathrm{s}}^2
 	=
 	\underline{\sigma}_0(t)+\underline{\sigma}(t),
 	\\
 \underline{\sigma}_0(t)
 	&=
 	\underline{X}(t)\underline{\sigma}_0\underline{X}^{\mathrm{T}}(t),
 	\\
 \underline{\sigma}(t)
 	&=
 	\int_0^t\mathrm{d}\tau\int_0^t\mathrm{d}\bar{\tau}~
 	\underline{X}(\tau)
 	\underline{\nu}(\tau-\bar{\tau})
 	\underline{X}^{\mathrm{T}}(\bar{\tau}),
 \end{align}
 \end{subequations}
 with {$\underline{\sigma}_0=\langle\Delta\mathbf{Q}^2(0)\rangle_{\mathrm{s}}$ as the covariance matrix} at $t=0$, $\underline{\nu}(\tau)=\mathrm{diag}[0,\nu(\tau)]$,
 $
 \underline{X}(t)
 $ as a matrix comprised of (time-derivates of) the response function
 (see Equation \eqref{Eq:DefX}),
 and the superscript ${\mathrm{T}}$ indicates the transpose of a matrix.

 It is interesting to note that, at arbitrary times, the autocorrelation of $\hat{q}$ is not a stationary function, i.e.,
 \begin{align}
 \label{Eq:Var1}
 \langle\hat{q}(t) & \hat{q}(t')\rangle_{\mathrm{s}}
 	\propto
 	\int_0^t\mathrm{d}x
 	\int_0^{t'}\mathrm{d}y~
 	\chi(x)\chi(y)
 	\nu(y-x+t-t'),
 \end{align}
 even though the response kernel $\chi$ and the correlator of the environment degrees of freedom $\nu$ are.
 Indeed, Equation \eqref{Eq:Var1} is given by a complex convolution of the system's self-consistent interaction with the environment. On top of that, the initial conditions evolve by construction, not necessarily in a stationary way.
 True stationarity can only be achieved at late times, where the impact of the initial conditions on the dynamics abates and the convolution can be expanded in Fourier modes of the form $\exp(-i\omega [t-t'])$ (see, e.g., Section III.B of ref. \cite{hsiang18} for details).
 Since the noise operator assumes a general Gaussian form, the statistics of the system are fully determined by the two-point function \cite{ford88}.
 We hence have that any even-order correlation reduces to the sum of two-point functions of all possible pairings of operators with preserved order, while any odd-order correlation vanishes \cite{ford65}. For example, we obtain for the fourth-order correlation function (see Appendix \ref{App:WicksTheorem} for a detailed proof).
 \begin{align}
 \label{Eq:HigherOrderCorr}
 \langle
 \hat{\xi}(t_1)\hat{\xi}(t_2)\hat{\xi}(t_3)\hat{\xi}(t_4)
 \rangle
   &=
   \langle
   \hat{\xi}(t_1)\hat{\xi}(t_2)
   \rangle
   \langle\hat{\xi}(t_3)\hat{\xi}(t_4)
   \rangle
 	\\\nonumber
   &+
   \langle
   \hat{\xi}(t_1)\hat{\xi}(t_3)
   \rangle
   \langle\hat{\xi}(t_2)\hat{\xi}(t_4)
   \rangle
 	\\\nonumber
 	&
   +
   \langle
   \hat{\xi}(t_1)\hat{\xi}(t_4)
   \rangle
   \langle\hat{\xi}(t_2)\hat{\xi}(t_3)
   \rangle.
 \end{align}
We note that the previous relation is not time-ordered and will come in handy when we compute the current fluctuations (see, e.g., Section \ref{Sec:TURHeat}).

 \subsection{Fluctuation--Dissipation Inequality and Robertson--Schr\"odinger Relation}\label{Sec:FDIRS}

Due to its strong formal resemblance to classical equations of motion, the quantum Langevin equation perhaps evokes the impression that the dynamics of the particle progresses deterministically: the particle absorbs energy from its environment via the ``force'' $\hat{\xi}$, processes it following its harmonic constraints, and emits it back into the environment; quantitatively described by the dissipation kernel $\gamma$.
 However, such a descriptions lets slide the quantum stochastic properties of the system.
 During the nonequilibrium evolution, even though the self-consistency of our approach ensures thermodynamic stability at all times, we are in lieu of an exact and transparent relation between the fluctuations of the environment [$\nu(\tau)$], the fluctuations of the reduced system of interest [$\langle\hat{q}(t)\hat{q}(t')\rangle_{\mathrm{s}}$], and the corresponding dissipation [$\gamma(\tau)$].
 Instead, even for systems which do not possess a fluctuation--dissipation relation, a corresponding \emph{fluctuation-dissipation inequality} (FDI) \cite{fleming13,reiche20a} can be found, and can serve as useful bounds on the physical quantities of the system evolving under dynamical conditions.
 In a recent work \cite{dong22}, we showed that a fluctuation--dissipation inequality exists at all times in the nonequilibrium dynamics of open quantum systems. We traced back the uncertainties of thermodynamic quantities in nonequilibrium systems to their quantum origins.
 We summarize the main points of the FDI in Appendix \ref{Sec:FDI}.
 For our purposes, it is sufficient to recall its main statement as \cite{fleming13,dong22}.
 \begin{align}
 \label{Eq:FDI}
 \int_0^t&\mathrm{d}\tau
 \int_0^t\mathrm{d}\tau'~
 f^*(\tau)
 \nu(\tau-\tau')
 f(\tau')
 \\\nonumber
 	&
 \geq
 	i\hbar
 	\int_0^t\mathrm{d}\tau
 	\int_0^t\mathrm{d}\tau'~
 	f^*(\tau)
 	\Gamma(\tau-\tau')
 	f(\tau')
 \end{align}
 for any complex function $f$.
 In particular, in a frequency domain, it is possible to show (see Appendix \ref{Sec:FDI})
 \begin{align}\label{Eq:FDIFrequ}
 \nu(\omega)\geq | i\hbar\Gamma(\omega)| = | \hbar\omega\gamma(\omega)|.
 \end{align}
 With this, it is straightforward to translate the results from the fluctuation--dissipation inequality for the environment to the fluctuations of the system of interest.
 Indeed, any property of $\nu$ and $\gamma$ is inherited by Equations \eqref{Eq:GreenFourier} and  \eqref{Eq:FluctSystem} such that also the fluctuations of the reduced system $\hat{q}$ are a positive semidefinite function.
 However, there is no such simple frequency--domain relation [Equation \eqref{Eq:FDIFrequ}] as in the case for the environment, since $\langle \hat{q}^2\rangle$ is not stationary [Equation \eqref{Eq:Var1}] at arbitrary times $t$.
 Alternatively, we can compare the position--momentum uncertainty prescribed by the fluctuation--dissipation inequality to the Robertson--Schrödinger inequality \cite{robertson29,schroedinger30}.
 In fact, it turns out that the former (FDI) can provide a stronger, more stringent bound to the uncertainties in the system at late times and only reduces to the Robertson--Schr\"odinger inequality by an additional application of the Cauchy--Schwarz inequality \cite{dong22}.
 The same situation is achieved for vanishingly small system+bath coupling, as one would expect from the limiting case of conventional thermodynamics \cite{hsiang18}.
 For details, we refer to Appendix \ref{Sec:RSI}.

 \section{Energy Flow and Entropy Production}
 \label{Sec:EnergyFlow}

 We are now interested in the direct consequences of the fluctuation--dissipation inequality on the system's thermodynamic properties.

 During the course of the interaction with the environment, the system's instantaneous mechanical Hamiltonian
 $
 \langle\hat{H}(t)\rangle_{\mathrm{s}}=\langle[\dot{\hat{q}}^2(t)+\omega_0^2\hat{q}^2(t)]\rangle_{\mathrm{s}}/(2)
 $
 varies with time starting from initially $\hbar\omega_0/2$, eventually increasing to its equilibrium value at late times.
 From the perspective of the system ($\hat{q}$), this is due to the (stochastic) interaction with the environment at any instance $t$ and can be associated to the power entering $P_{\mathrm{in}}$ and exiting $P_{\mathrm{out}}$ the system, respectively.
 Upon multiplying from both sides of the quantum Langevin equation (Equation \eqref{Eq:QLE0}) with $\dot{\hat{q}}$, we obtain
 \begin{subequations}
 \begin{align}
 \label{Eq:PowerDef}
 P_{\mathrm{in}}(t)
 	&=
   \langle
 	\hat{\xi}(t)
 	\dot{\hat{q}}(t)
 	\rangle_{\mathrm{s}}
 	=
 	\int_0^t\mathrm{d}x~
 	\dot{\chi}(x)
 	\nu(x)
 	,
 	\\
 \label{Eq:PowerDef2}
 P_{\mathrm{out}}(t)
 	&=
   2
 	\int_0^t\mathrm{d}\tau~
 	\Gamma(t-\tau)
   \left\langle
 	\hat{q}(\tau)
 	\dot{\hat{q}}(t)
   \right\rangle_{\mathrm{s}}.
 \end{align}
 \end{subequations}
 We remark that the incoming power is unaffected by the initial conditions.
 The outgoing power, on the other hand, is determined by the cross-correlation of position and momentum of the system, and hence by both the initial conditions as well as the response to the noisy environment (see Equation \eqref{Eq:qSol} for the exact expression).
 Due to the linearity of our system+bath coupling and the fact that $\langle\hat{q}(0)\hat{\xi}\rangle_{\rm s}=\langle\dot{\hat{q}}(0)\hat{\xi}\rangle_{\rm s}=0$, the corresponding dynamics decouple, and we can write
 $
 P_{\mathrm{out}}(t)
   =
   P_{\mathrm{out}}^{\mathrm{init}}(t)
   +
   P_{\mathrm{out}}^{\mathrm{fluc}}(t)
 $
 (see also Equation \eqref{Eq:FluctSystem}).
 As one could have expected, the information and energy initially stored in the system dissipates into the environment over time.
 Instead, over time, additional excitations are transferred to the system from the fluctuating environment.
 These environment-induced excitations again become dissipated.
 The system and environment are self-consistently back-acting onto one another.
 At late times, the system can reach a nonequilibrium steady-state (equilibrium in our case \cite{ford88a,hsiang18}).
 Concretely, upon defining the part of the system's dynamics, denoted by $\hat{\mathcal{Q}}(t)$, due to the noise operator $\xi$ (Equation \eqref{Eq:qSol}).
 \begin{subequations}
 \label{Eq:qFluc}
 \begin{align}
 \hat{\mathcal{Q}}(t)
   &=\int_0^t\mathrm{d}\tau~\chi(t-\tau)\hat{\xi}(\tau)
 ,
 \end{align}
 \end{subequations}
 The change in the system's instantaneous mechanical energy due to the fluctuating interaction with the environment obeys the relation (see Appendix \ref{App:CalcFlucEnergyFlow} for details).
 \begin{align}
 \label{Eq:powerFluct}
 &\frac{\mathrm{d}}{\mathrm{d}t}
 	\left\langle
 	\frac{1}{2}\dot{\hat{\mathcal{Q}}}^2(t)
 	+
 	\frac{\omega_0^2}{2}\hat{\mathcal{Q}}^2(t)
 	\right\rangle_{\mathrm{s}}
 	=
 	\langle
 	\dot{\hat{\mathcal{Q}}}\ddot{\hat{\mathcal{Q}}}
 	+
 	\omega_0^2\hat{\mathcal{Q}}\dot{\hat{\mathcal{Q}}}
 	\rangle_{\mathrm{s}}
 	\\\nonumber
 	&=
 	\int_0^t\mathrm{d}x~
 	\dot{\chi}(x)
   \int_0^t\mathrm{d}z~
   \left(
   \delta(z)
   +
 	\left[
 	\ddot{\chi}(z)
 	+\omega_0^2
 	\chi(z)
 	\right]
   \right)
   \nu(x-z)
 	\\\nonumber
 	&=
 	P_{\mathrm{in}}(t)-P_{\mathrm{out}}^{\mathrm{fluc}}(t)
   \to 0, \quad t\to\infty.
 \end{align}
 At late times, we can further invoke the fluctuation--dissipation inequality to derive a lower bound on the fluctuating part of the, say, outgoing power.
 Indeed, upon partially integrating with respect to $\tau$ in Equation \eqref{Eq:PowerDef2}, we can write
 \begin{align}
 \label{Eq:PoutTimeDomain}
 P_{\mathrm{out}}^{\mathrm{fluc}}
   &=
   2
 	\int_0^t\mathrm{d}\tau~
 	\gamma(t-\tau)
   \left\langle
 	\dot{\hat{\mathcal{Q}}}(\tau)
 	\dot{\hat{\mathcal{Q}}}(t)
   \right\rangle_{\mathrm{s}}
   \\\nonumber
   &=
   \int_0^t\mathrm{d}\tau~
   \gamma(t-\tau)
   \\\nonumber
   &
   \times
   \int_0^{\tau}\mathrm{d}x
   \int_0^t\mathrm{d}y~
   \dot{\chi}(x)
   \dot{\chi}(y)
   \nu(y-x+\tau-t).
 \end{align}
 Performing the limit $t\to\infty$, we obtain
 \begin{align}
 \label{Eq:POutFluc}
 P_{\mathrm{out}}^{\mathrm{fluc}}
 &\to
 \lim_{t\to\infty}
 \int_0^{t}
 \mathrm{d}x
 ~
 \gamma(x)
 \int\frac{\mathrm{d}\omega}{2\pi}
 \omega^2
 \alpha(\omega)\nu(\omega)\alpha^*(\omega)
 e^{i\omega x}
 \\\nonumber
 &=
 \int\frac{\mathrm{d}\omega}{2\pi}
 \omega^2
 \gamma_{\theta}(\omega)
 |\alpha(\omega)|^2\nu(\omega)
 \\\nonumber
 &\geq
 \hbar
 \int\frac{\mathrm{d}\omega}{2\pi}~
 \gamma_{\theta}(\omega)
 |\omega^3\gamma(\omega)|
 |\alpha(\omega)|^2,
 \end{align}
 while $P_{\mathrm{out}}^{\mathrm{init}}\to 0$.
 Since they equal at late times, the same relation holds for $P_{\mathrm{in}}$.
 Clearly, the net flow of energy into the system due to the interaction with the environment modifies its mean energy, and hence leads to a change in the particle's von Neumann entropy over the course of equilibration.

 Due to the linearity of our system, the density matrix can be calculated exactly (see Appendix \ref{App:DensityMatrix} and references therein).
 Quite remarkably, the result shows a formal resemblance to the density matrix of a quantum harmonic oscillator coupled to a bath with \emph{{time-dependent}
 } temperature \cite{pagel13,hsiang21}.
 In particular, the entropy of the system at time $t$ is given by \cite{hsiang21}.
 \begin{align}
 \label{Eq:Entropy}
 &\mathcal{S}(t)
   =
   \frac{\hbar\omega_0}{e^{\hbar\omega_0\beta(t)}-1}
   -\log\left[
   1-e^{-\hbar\omega_0\beta(t)}
   \right]
   \\\nonumber
   &=
   \left(\sqrt{u}+\frac{1}{2}\right)
   \log\left[\sqrt{u}+\frac{1}{2}\right]
   -
   \left(\sqrt{u}-\frac{1}{2}\right)
   \log\left[\sqrt{u}-\frac{1}{2}\right]
 \end{align}
 with the effective parameter $\beta(t)$ (see Equation \eqref{Eq:BetaEff} of Appendix \ref{App:DensityMatrix}) and $u=u(t)=\det[\underline{\sigma}_0(t)+\underline{\sigma}(t)]/\hbar^2$.
 One can readily check that this is equivalent to the von Neumann entropy
 $\mathcal{S}=\mathcal{S}_{\rm vN}=-\text{Tr}[\hat{\rho}\log\hat{\rho}]$, where the trace is performed over the system's degrees of freedom, see, e.g., refs. \cite{joos85,koks97,colla21}.
 From the perspective of the system, it is further interesting to note that the von Neumann entropy production rate is bounded from below by the time-derivative of its covariance matrix.
 \begin{align}
 \label{Eq:BoundEntropyProduction}
 \partial_t\mathcal{S}
 &=
 \text{arccsch}\left[\frac{\sqrt{u(t)-\frac{1}{4}}}{2}\right]
 \frac{\dot{u}(t)}{\sqrt{u(t)}}
 \\\nonumber
 &
 \geq
 \frac{\dot{u}(t)}{2u(t)}
 =
 \partial_t \log\left[\sqrt{u(t)}\right]
 .
 \end{align}
 Starting with the contact between system and environment at $t=0$ to the equilibration at $t\to\infty$, the von Neumann entropy varies.
 The total production of entropy in the system is given by
 \begin{align}
 \Delta\mathcal{S}\equiv \mathcal{S}(t\to\infty)-\mathcal{S}(0).
 \end{align}
 We note that this is not the total entropy production of the combined system, which would also need to consider the heat flow between system and environment (see, e.g., refs. \cite{pucci13,colla21} for a comparison of different definitions for the total entropy).
 At the beginning of the experiment, the system and environment are decoupled.
 If we further assume vanishing means and an unsqueezed initial state with $\underline{\sigma}(0)=\text{diag}[1/2,1/2]$, the system is in its well-defined ground state, hence $\mathcal{S}(0)=0$.
 At late times, for comparison, the interaction with the environment has increased the uncertainty in the state of the system with respect to its initial (isolated) value, and we have that the entropy reaches its {equilibrium} value.
 Equation~\eqref{Eq:Entropy} is a monotonic function of $u$, but not necessarily of time $t$, as $u$ is not a monotonic function of time.
 For a discussion of the von Neumann entropy over the course of the nonequilibrium evolution, we refer to ref. \cite{dong22}.
 At late times, however, it becomes a constant depending on the system parameters as well as the coupling strength.
 For instance, in the case of strong system+bath coupling, where $u\gg 1/4$, we have that
 $\Delta\mathcal{S}\sim 1+\log[u]$ {varying} logarithmically with $u$.
 Hence, the uncertainty relations derived in Equation \eqref{Eq:FDIsConnected} (see also Equation \eqref{Eq:POutFluc}) are inherited also by $\Delta\mathcal{S}$ in exactly the same hierarchy.
 In this way, the fluctuation-dissipation inequality provides a lower bound to the entropy production in the system. Incidentally, that bound exceeds the bound one would expect on the basis of the Robertson--Schrödinger inequality.
 By implication, similar arguments can be drawn for the effective temperature $\beta(t)^{-1}$ at late times.

\begin{figure}[h]
\centering
\includegraphics[width=0.48\textwidth]{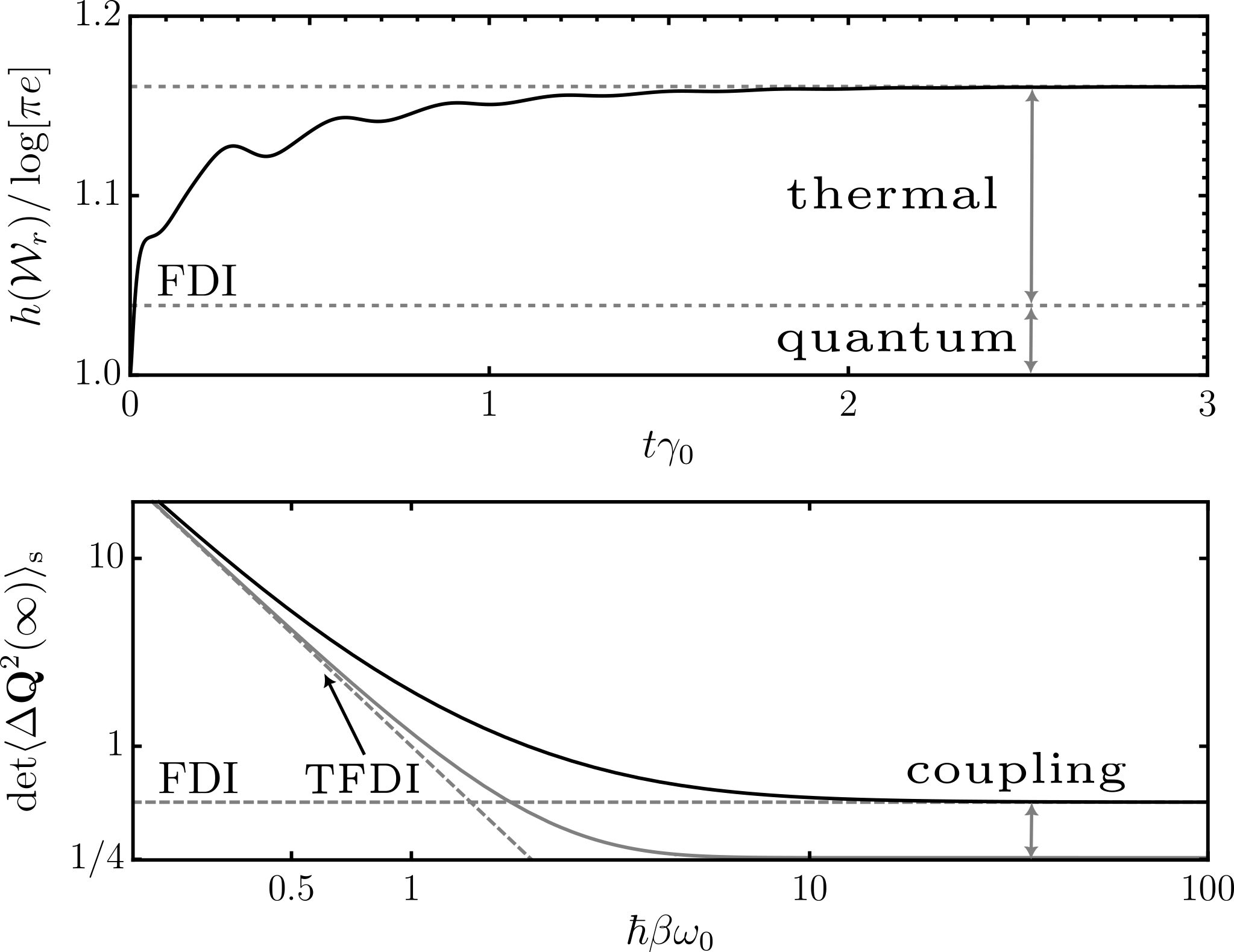}
\caption{
(\textbf{top})
 Phase space entropy for positive Wigner functions (Equation \eqref{Eq:hWr}, normalized to the minimal uncertainty $\log[\pi e]$) as a function of time (measured in multiples of the dissipation rate).
 We employ the bath spectral density of Equation \eqref{Eq:ILambda} and use parameters $\omega_0/\gamma_0=10$, $\Lambda/\omega_0=10$, and dimensions where $\hbar=1$ as well as choose $\underline{\sigma}_0=\text{diag}[1/2,1/2]$ for the initial conditions.
 The lower dashed line gives the lower bound prescribed by the fluctuation-dissipation inequality (FDI) at late times (Equations \eqref{Eq:FDI}, \eqref{Eq:FDIFrequ} and \eqref{Eq:hWr}), which can be connected to quantum fluctuations in the coupled system+bath system.
 The upper dashed line gives the exact late time limit of the Gaussian evolution (Equations \eqref{Eq:hWr} and \eqref{Eq:RSUncertainty3}), which also includes thermal fluctuations (see Section \ref{Sec:ThermalFDI}).
 (\textbf{bottom})
 Late-time quantum uncertainty [Equations \eqref{Eq:FluctSystem} and \eqref{Eq:RSUncertaintyFDI}] as a function of $\hbar\beta\omega_0$, i.e., a measure of the respective impact of quantum or thermal fluctuations.
 For finite system-bath coupling (black, solid line; $\omega_0/\gamma_0=1$), the uncertainty always exceeds the minimal bound of $1/4$ given by the Robertson-Schr\"odinger equation and saturates the fluctuation-dissipation inequality for $\hbar\beta\omega_0\gg1$ (gray, horizontal, dashed line).
 This discrepancy fades for smaller coupling  (gray, solid line; $\omega_0/\gamma_0=100$).
 Additionally, for $\hbar\beta\omega_0\ll1$, thermal fluctuations start to prevail over the quantum fluctuations, and the more accurate bound (comparing to the FDI) can be provided by the thermal fluctuation-dissipation inequality (TFDI; gray, dashed, nonhorizontal line; see Section \ref{Sec:ThermalFDI}).}
\label{Fig:UncertaintyPlot}
 \end{figure}

 Lastly, we comment on the consequences of the fluctuation--dissipation inequality on the Shannon entropy in the phase--space (Wigner) representation, i.e., the \emph{entropic uncertainty relation
 } \cite{bialynickibirula75,coles17}.
 For nonvanishing second-order cross-correlation and Gaussian systems with \emph{{positive}
 } Wigner function, the entropic uncertainty relation can be formulated on the basis of the system's Wigner function \cite{hertz17,vanherstraeten21}.
 \begin{align}
 \label{Eq:hWr}
 h(\mathcal{W}_r)
   &:=
   -\int\mathrm{d}q\int\mathrm{d}\dot{q}~
   \mathcal{W}_r(q,\dot{q},t)
   \log\left[\mathcal{W}_r(q,\dot{q},t)\right]
   \\\nonumber
   &
   =
   \log[e\pi]
   +
   \log\left[2\sqrt{u(t)}\right],
 \end{align}
 where, in the second line, we inserted the expression for the Wigner function in Equation \eqref{Eq:WignerFunction} and performed the Gaussian integrals.
 Equation \eqref{Eq:hWr} gives the Shannon entropy of the Wigner distribution and can be particularly useful from the information theoretical point of view in multipartite Gaussian systems \cite{adesso12}.
 Even though the difference is rather subtle in our single-oscillator case, we remark that $h(\mathcal{W}_r)$ is not primarily connected to the von Neumann entropy (i.e., a R\'enyi-1 entropy $\mathcal{S}$) but rather the generalized concept of the R\'enyi-2 entropy $\mathcal{S}_2=-\text{log}\text{Tr}(\hat{\rho}_S^2)=h(\mathcal{W}_r)-\log[e\pi]$ (compare also to Equation \eqref{Eq:BoundEntropyProduction}).
 We refer to \cite{adesso12} and references therein for further details.
In Figure \ref{Fig:UncertaintyPlot}, we report a numerical evaluation of the uncertainty and the (positive) Wigner entropy over the course of the nonequilibrium evolution as well as study the distinct impact of quantum and thermal fluctuations.
Clearly, from the Robertson-Schr\"odinger inequality, $\sqrt{u(t)}\geq 1/2$, one has $h(\mathcal{W}_r(t))\geq\log[e\pi]$, which is the traditional statement of the entropic uncertainty relation.
For instance, in our case, at $t=0$, where $\sqrt{u(0)}=1/2$ corresponds to a pure Gaussian state, the inequality is saturated, $h(\mathcal{W}_r(0))=\log[e\pi]$.
At finite times, however, even though the system's Wigner function remains Gaussian, the system+bath coupling introduces additional fluctuations into the system so that the Wigner entropy $h(\mathcal{W}_r)$ exceeds its lower bound.
Here, for comparison, the fluctuation--dissipation inequality in Equation \eqref{Eq:RSUncertaintyFDI} can provide a more precise statement. Indeed, since it is equal to or exceeds the conventional Robertson-Schr\"odinger relation [Equation \eqref{Eq:FDIsConnected}], the fluctuation--dissipation inequality can be used to extract information on the system+bath coupling from the entropic uncertainty relation, even when no exact solution (as is the case for Gaussian systems) is available.

 \section{Quantum Thermodynamic Uncertainty Relation}\label{Sec:TUR}

 \subsection{Thermal Fluctuation--Dissipation Inequality}\label{Sec:ThermalFDI}

 So far, our results are derived from mathematical and conceptual considerations, i.e. the hermicity of the noise operator and the self-consistency of the interaction.
 We now turn our full attention to the specific case of quantum Brownian motion and derive a thermodynamic uncertainty relation for the nonequilibrium energy current between system and environment.
 To this end, we write the dissipation and the noise kernel in the convenient form (see, e.g., refs. \cite{caldeira83a,grabert87}).
 \begin{subequations}
 \label{Eq:NuGamma}
 \begin{align}
 \gamma(\tau)
 	&
   =
   \frac{1}{2}
   \int\mathrm{d}\omega~
 	\frac{I(\omega)}{\omega}
   e^{-i\omega\tau},
 	\\
 	\label{Eq:nuTimeQBM}
 \nu(\tau)
 	&=
 	\frac{\hbar}{2}
   \int\mathrm{d}\omega~
 	I(\omega)\coth\left[\frac{\hbar\beta\omega}{2}\right]
   e^{-i\omega\tau},
 \end{align}
 \end{subequations}
 where $I(\omega)$ is the bath spectral density and $\beta^{-1}=k_{\rm B}T$ the temperature of the environment with $k_{\rm B}$ the Boltzmann constant (not to be confused with the effective parameter used in Equation \eqref{Eq:Entropy}).
 The bath spectral density is an odd function in frequency.
 Moreover, using the relation
 $
 \omega\gamma(\omega)
   =
   2\text{Im}[\Gamma_{\theta}(\omega)]
 $, we have that
 \begin{align}
 \label{Eq:SpectralDen}
 I(\omega)
   &=
   \frac{2}{\pi}
   \text{Im}
   [\Gamma_{\theta}(\omega)].
 \end{align}
 $\Gamma_{\theta}(\omega)$ is an analytic function in the upper complex frequency plane with singularities, i.e., the physical resonances of the system, that are symmetrically distributed with respect to the complex frequency axis.
 For each singularity at $\omega_n$ with $\text{Im}[\omega_n]<0$, solving $\Gamma_{\theta}^{-1}(\omega_n)=0$, $\Gamma_{\theta}(\omega)$ also features a singularity at $-\omega_n^*$ (usually referred to as crossing relation).
 We note that this allows for (degenerate) purely complex solutions.
 Hence, $\omega\gamma(\omega)$ is in general \emph{{not}
 } analytic in the complete complex frequency plane.
 Instead, due to the $\text{Im}[\cdot]$ in Equation~\eqref{Eq:SpectralDen}, for each singularity at $\omega_n$, the spectral density $I(\omega)$  shows the pair $\{\omega_n,\omega_n^*\}$ as poles in the complex frequency plane.
 In Appendix \ref{App:BathSpectrDens}, we summarize the properties of dissipation kernels for some common bath spectrum models.

 Due to the simple form of Equation \eqref{Eq:NuGamma}, a Fourier transform is well-defined at all times, and we obtain the \emph{{exact}
 } relation.
 \begin{align}
 \label{Eq:FDT}
 \nu(\omega)
   =
   \hbar\omega
   \coth\left[\frac{\hbar\beta\omega}{2}\right]\gamma(\omega)
   \geq |\hbar\omega\gamma(\omega)|.
 \end{align}
 This is the celebrated (local) fluctuation--dissipation theorem \cite{callen51,haenggi05} and means that for the bath degrees of freedom in quantum Brownian motion, detailed balance is fulfilled at all times.
 We note that the latter \emph{{equality}
 } in Equation \eqref{Eq:FDT} follows as a special case for zero temperature since
 $
 \coth{\hbar\beta\omega/2}\to\text{sgn}[\omega]
 $
 for $\beta\to\infty$.
 Indeed, for vanishing temperatures, the equality holds even in time domain.
 \begin{align}
 \label{Eq:ZeroTempCanonicalFDI}
 &\nu(\tau)
   =-i\hbar\partial_{\tau}\gamma(\tau)
   =i\hbar\Gamma(\tau),
 &\beta\to\infty.
 \end{align}
 The difference per frequency between $\nu$ and $\gamma$ stems from the thermal occupation of the field.
 Moreover, using that $\coth[x\omega]\geq [x\omega]^{-1}$ ($\omega>0$), we can relate in frequency domain
 $\nu(\omega)\geq (2/\beta)\gamma(\omega)$.
 The {corresponding} relation {in time domain} does not generally hold since the Fourier transform [Equation \eqref{Eq:NuGamma}] does not necessarily preserve order.
 Instead, we can specify the notion of the fluctuation--dissipation inequality in Equation \eqref{Eq:FDI} by using our knowledge of the detailed balance of the bath degrees of freedom in quantum Brownian motion, i.e.,
 \begin{align}
 \label{Eq:ThermalInequality}
 &\int_0^t\mathrm{d}\tau
 \int_0^t\mathrm{d}\tau'
 f(\tau)
 \nu(\tau-\tau')
 f^*(\tau')
 \\\nonumber
   &
 \geq
   \int_0^t\mathrm{d}\tau
   \int_0^t\mathrm{d}\tau'
   f(\tau)
   f^*(\tau')
   \begin{cases}
   \frac{2}{\beta}\gamma(\tau-\tau'), &\beta^{-1}\ll\hbar\omega_c,
   \\
   i\hbar \Gamma(\tau-\tau'), & \forall \beta.
   \end{cases}
 \end{align}
 For brevity, and in order to discern Equations \eqref{Eq:FDI} {(second line of the previous equation)} and \eqref{Eq:ThermalInequality}, we  denote the {first line of the previous equation} as the \emph{{thermal}
 } fluctuation--dissipation inequality.
 We emphasize that the first line of Equation \eqref{Eq:ThermalInequality} is not just a high-temperature approximation of the second line.
 When comparing with Equation \eqref{Eq:FDI}, the thermal version of Equation \eqref{Eq:ThermalInequality} is valid for temperatures larger than the resonances of the system+environment composite.
 It provides a more accurate estimate than \eqref{Eq:FDI} does in the classical, high-temperature regime.
 In this regime, it has the advantage of providing the tighter lower bound on the system's fluctuations.
 The quantitative advantage, however, comes at the price of weakened generality.
 Indeed, at lower temperature, it needs to be replaced by the inequality in Equation \eqref{Eq:FDI} which gives the absolute quantum limit.
 We finish this paragraph by considering some concrete examples for the thermal fluctuation--dissipation inequality as well as deduce its consequences on the fluctuations of the system degrees of freedom.

 While the fluctuation--dissipation inequality [Equation \eqref{Eq:FDT}] gives the absolute quantum limit of the fluctuation's correlations and provides the most accurate bound for large frequencies (small time delays $\tau$), the thermal inequality (Equation \eqref{Eq:ThermalInequality}) puts the focus on the environment's temperature and becomes more accurate at small frequencies (large time delays $\tau$). {As expected, the thermal fluctuation--dissipation inequality is hence connected to the classical thermodynamic limit.}
 For analytical details on how the noise kernel is affected by different types of bath spectral densities, we refer to Appendix \ref{App:IntNoiseKernel}.

 \subsection{Numerical Results and Quantifying Error}

\begin{figure}[h]
     \centering
     \includegraphics[width=0.48\textwidth]{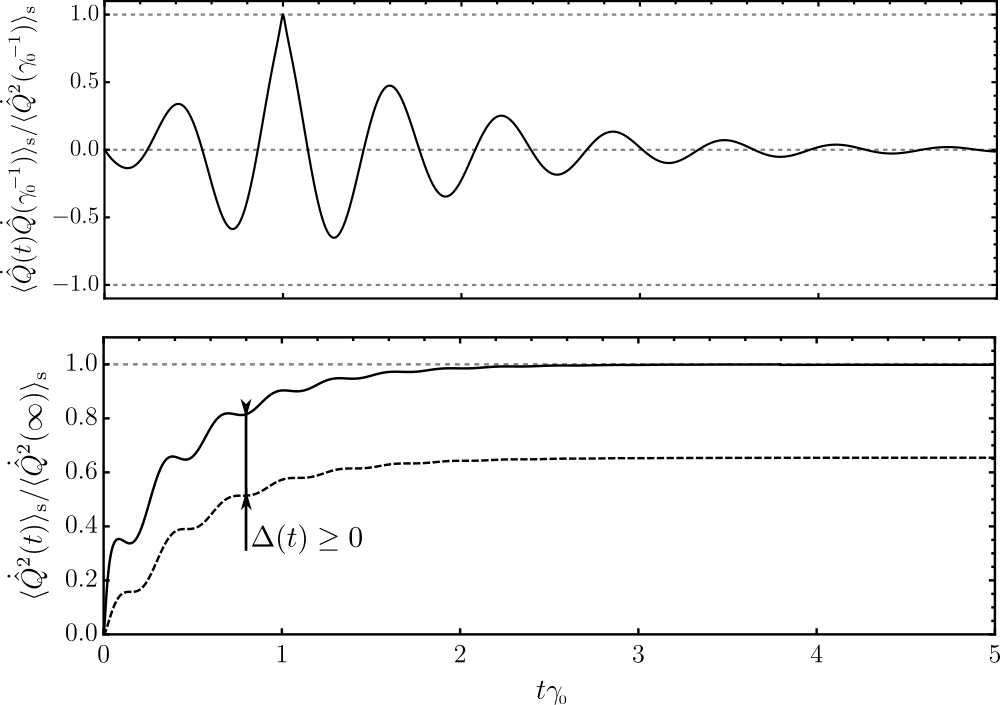}
     \caption{Numerical
  evaluation of the symmetric fluctuations $\langle\dot{\hat{\mathcal{Q}}}(t)\dot{\hat{\mathcal{Q}}}(t')\rangle_{\rm s}$ of the system's momentum operator solely connected to the fluctuating dynamics (see Equation \eqref{Eq:qFluc}) as a function of time in multiples of the dissipation rate $\gamma_0$.
     We employ the $\Lambda$-model in Equation \eqref{Eq:ILambda} for the bath spectral density and use parameters $\omega_0/\gamma_0=10$, $\Lambda/\omega_0=10$, and$\hbar\beta\omega_0=2$ and work in dimensionless units where $\hbar=1$.
     {(\textbf{top})}
     Two-time correlation centered at the dissipation time $\gamma_0^{-1}$ and normalized to its equal-time correlation at $t=\gamma_0^{-1}$.
     At $t=\gamma_0^{-1}$, the apparent kink is owed to the numerical resolution in time. The curve is smooth.
     {(\textbf{bottom})}
     Equal-time correlation normalized to its late-time limit (solid).
     Lower bound prescribed by thermal fluctuation--dissipation inequality (dashed).
     The difference between the two is given by $\Delta(t)$ (Equation \eqref{Eq:DeltaqDot}).
     }
     \label{fig:qdot}
 \end{figure}

 We report the numerical results for the autocorrelation of the momentum operator using the $\Lambda$-model for the bath spectral density (see Equation \eqref{Eq:ILambda} and Appendix \ref{App:IntNoiseKernel} for details on the numerics), i.e., $\langle\dot{\hat{\mathcal{Q}}}(t)\dot{\hat{\mathcal{Q}}}(t')\rangle_{\rm s}$, in Figure \ref{fig:qdot}.
 We take the momentum correlation to illustrate our findings as they  play a role in the next section dealing with the energy flow between system and environment.
 As we anticipated from the expressions for the noise correlations $\nu$ and the response function $\chi$, the autocorrelation is maximal if $t=t'$ and decays exponentially for increasing time delay $|t-t'|$ (top of Figure \ref{fig:qdot}).
 Per construction (see Equation \eqref{Eq:qFluc}), the momentum correlator vanishes if either $t$ or $t'$ is smaller than zero, and the asymmetric shape with respect to the value at $t=t'$ stems from the fact that we have set $t_i=0$ as the initial time of our experiment. For comparison, if we were to set $t_i\to-\infty$, the shape would be fully symmetric.
 The momentum correlation at equal times increases over time and reaches its late-time limit for times $t\gg1/\gamma_0$ (see Equation \eqref{Eq:RSUncertainty3}) (bottom of Figure \ref{fig:qdot}).
 As expected from Equation \eqref{Eq:ThermalInequality}, at all times, the momentum correlation exceeds the value prescribed by the thermal fluctuation--dissipation inequality, where we replace $\nu\to(2/\beta)\gamma$ in the evaluation of the time integrals (dashed line, bottom of Figure \ref{fig:qdot}).

 Starting from the microscopic model in Equation \eqref{Eq:NuGamma}, the difference between a regime that is dominated by quantum fluctuations and a regime that is dominated by thermal fluctuations becomes immediately clear from the coth function weighted by the bath spectral density.
 From this perspective, the thermal fluctuation--dissipation inequality in Equation \eqref{Eq:ThermalInequality} may appear deceptively simple.
 However, when concrete examples for realistic situations are considered, where the coth function and the bath spectral density are buried in many layers of modeling and experimental circumstances, the situation can be less clear.
 Sometimes, it may seem legitimate to neglect certain types of fluctuations in the interaction, and hence it becomes interesting to quantify the resulting error.
 To this end, let us consider the difference between the exact momentum fluctuations and their lower limit in terms of the thermal fluctuation--dissipation inequality, i.e.,
 \begin{align}
 \label{Eq:DeltaqDot}
 \Delta(t)
   &:=
 \int_0^t\mathrm{d}x\int_0^t\mathrm{d}y~
 \dot{\chi}(x)\dot{\chi}(y)
 \left[
 \nu(y-x)-\frac{2}{\beta}\gamma(y-x)
 \right]
 \\\nonumber
 &=
 \int_0^t\mathrm{d}x\int_0^t\mathrm{d}y
 \int\mathrm{d}\omega~
 e^{-i\omega(y-x)}
 \frac{\dot{\chi}(x)\dot{\chi}(y)}{2}
 \\\nonumber
 &\quad\times
 \frac{2}{\beta}
 \left[
 \frac{\hbar\beta\omega}{2}
 \coth\left[\frac{\hbar\beta\omega}{2}\right]
 -1
 \right]
 \gamma(\omega),
 \end{align}
 which is strictly positive by means of Equation \eqref{Eq:ThermalInequality}.
 The previous relation reveals quite intuitively the physical meaning of the thermal fluctuation--dissipation inequality.
 The fluctuations of the system can be connected to their thermal ($\sim$$\beta^{-1}$) and quantum ($\sim$$\hbar\omega/2$) nature.
 Both are interwoven ($\sim$$\coth[\hbar\beta\omega/2]$) into the non-Markovian and nonequilibrium dynamics of the evolving system by means of the time and frequency integrals in Equation~\eqref{Eq:DeltaqDot}.
 Furthermore, they are weighted by the properties of the system and the environment ($\chi$ and $\gamma$) which set the relevant energy scales for the quantum fluctuations, i.e., $\omega_0$ and $\Lambda,\gamma_0$ or $\Omega,\kappa,\rho(\mathbf{r}_a)$ for the model in Equation \eqref{Eq:ILambda} or \eqref{Eq:SpecDenOscillator} (see Appendix \ref{App:BathSpectrDens}), respectively.
 The thermal fluctuation--dissipation inequality now puts a special emphasis on the thermal (classical) fluctuations of the interaction, i.e., approximates the hyperbolic cotangent in the frequency domain (Equation \eqref{Eq:nuTimeQBM}) by its value for small arguments.
 Since there is no way to circumvent quantum mechanics, the corresponding fluctuations  always come on top, and hence the thermal fluctuation--dissipation inequality provides a lower bound for sufficiently high temperatures (see discussion after Equation \eqref{Eq:ThermalInequality}).
 As a consequence, in the extreme limit of vanishing temperature, the thermal fluctuation--dissipation inequality becomes trivial and the general fluctuation--dissipation inequality or the Robertson--Schrödinger inequality (see, e.g., Equation \eqref{Eq:FDIsConnected}) provide a sharper and physically nontrivial bound, since their focus lies on the quantumness of the system.
 Even though such considerations are well-known from equilibrium physics or nonequilibrium steady states, our considerations show that such concepts can to some extent be translated into the full nonequilibrium evolution of the system.
 In other words, Equation \eqref{Eq:DeltaqDot} could be understood as a measurement of the ``quantumness'' of the nonequilibrium dynamics of a quantum Brownian particle that also takes the particular bath spectral density into account.
 Indeed, for illustration, let us consider the Markovian and late-time (equilibrium) limit of Equation \eqref{Eq:DeltaqDot}.
 Using the $\Lambda$ model (Equation~\eqref{Eq:ILambda} and Appendix \ref{App:BathSpectrDens}) in the limit $\Lambda\to\infty$ (formally yields the Markovian limit) and additionally assuming that $\gamma_0\ll\omega_0$, the polarizability $|\alpha(\omega)|^2\sim \pi\delta(\omega^2-\omega_0^2)/(2\gamma_0\omega)$ and we obtain
 $
 \lim_{t\to\infty}
 \Delta(t)
   =
   ([\hbar\beta\omega/2]\coth[\hbar\beta\omega/2]-1)/\beta
   \to \hbar\omega_0/2
 $ as
 $\beta\to\infty$.
 The fluctuations of the (momentum) operator are purely quantum.
 For a numerical evaluation of the role of the thermal fluctuation--dissipation inequality on the uncertainty function of the system, we refer to Figure~\ref{Fig:UncertaintyPlot}.

 For a more advanced example, we can extend the previous discussion to the interaction of an atom with the material-modified electromagnetic field.
 Considering alkali-metal atoms and conducting macroscopic bodies, thermal fluctuations are usually much smaller than any of the system's resonances \cite{dalvit11}.
 If the atom moves with constant velocity in the vicinity of a macroscopic body, it experiences a decelerating force, the so-called quantum friction \cite{pendry97}. Even at zero temperature, the motion-induced Doppler shift then instigates additional low-frequency fluctuations into the system that, for simplicity, can be understood as mimicking certain aspects of thermal fluctuations \cite{intravaia16b,reiche21}.
 The important point is that, if one assumes that equilibrium can be established locally, even though the system is in a nonequilibrium state \cite{intravaia16a},
 the power incoming into the system can be written in a form very similar to Equation \eqref{Eq:DeltaqDot} (see Equation (11) in ref. \cite{reiche20a}).
 In the particular example, a nonvanishing $\Delta(t)$ hinted towards the negligence of important fluctuations in the power spectrum \cite{reiche21} and could be shown to be a serious defect in the underlying statistical modeling of the interaction \cite{reiche20a}.
 One is hence often well-advised to minimize $\Delta(t)$ in (quantum) fluctuation-induced systems.


 \section{Nonequilibrium Current, Energy Flow, and Current Fluctuations}\label{Sec:TURHeat}

\begin{figure}[t]
     \centering
     \includegraphics[width=0.48\textwidth]{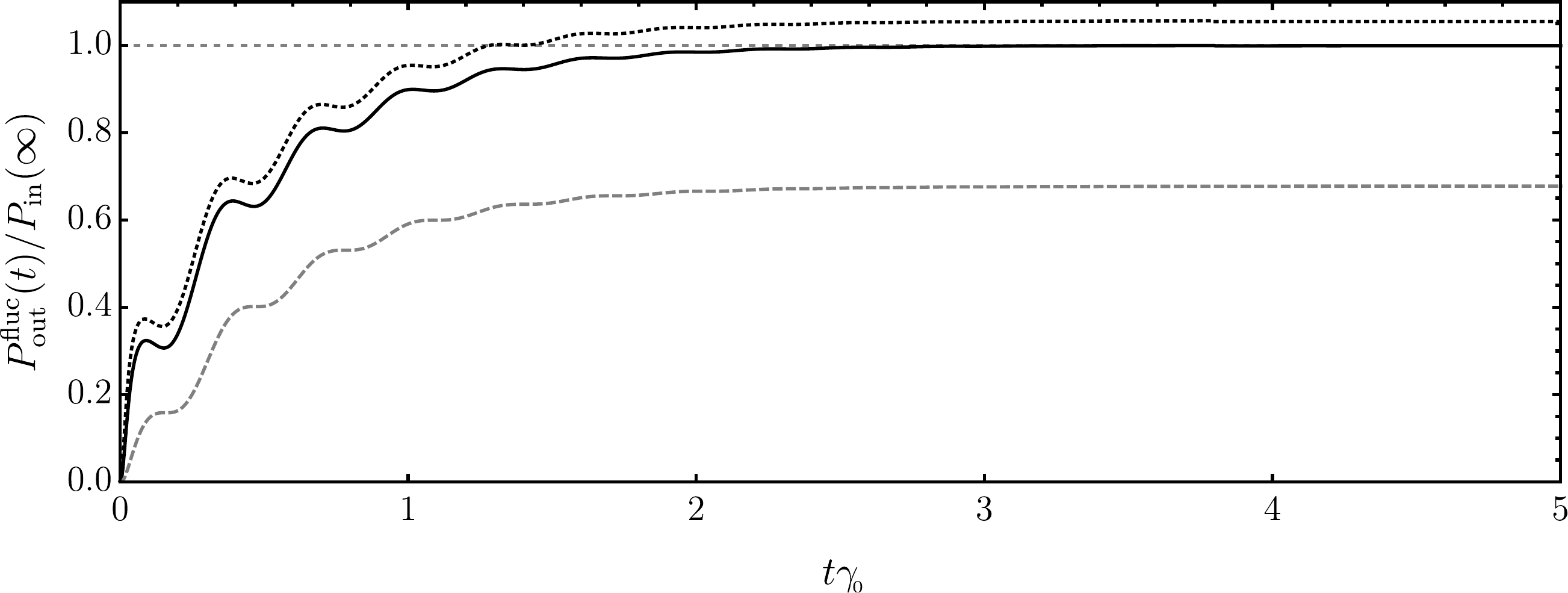}
     \caption{Outgoing power connected to the fluctuating dynamics of the system $P_{\rm out}^{\rm fluc}$ (Equations \eqref{Eq:PoutTimeDomain} and \eqref{Eq:JTotalPower}) as a function of time in multiples of the dissipation rate $\gamma_0^{-1}$.
     Parameters are chosen as in Figure \ref{fig:qdot}.
     We normalize to the expression for the ingoing power at late times $P_{\rm in}(\infty)$ [Equation~\eqref{Eq:PowerDef}] in order to indicate the balancing of the two at equilibrium (solid line).
     We further report the corresponding expression using the thermal fluctuation--dissipation inequality, i.e., replacing $\nu\to(2/\beta)\gamma$ in the numerical evaluation, which is always smaller than the full expression (dashed).
     Lastly, we give the upper estimate of Equation \eqref{Eq:PowerOutUpperLimit} (dotted).
     \label{fig:mesh1}}
 \end{figure}

 In the literature on thermodynamic uncertainty relations, it has become customary to use a nonequilibrium current operator to quantify the degree of nonequilibrium, namely, the
 thermodynamic uncertainty relations provide a lower bound on the fluctuations of that current operator \cite{horowitz20}.
 Quite generally, we can define the current operator as
 \begin{align}
 \label{Eq:Current}
 \hat{J}(t)
   :=
   \frac{1}{2}
   \int_0^t
   \mathrm{d}\tau~
 	\left\{
   \hat{f}(\hat{q}(\tau),\dot{\hat{q}}(\tau),\hat{\xi})
 	,
 	\dot{\hat{q}}(\tau)
 	\right\}
 \end{align}
 with an, in principle, arbitrary function $f$ that, for simplicity, is not explicitly time-dependent.
 We use the anticommutator here in order to ensure the real expectation values in our later examples.
 Various examples of such  can be found in the literature (see Section \ref{Sec:Introduction}).
 As the particle is constantly exchanging energy with its surroundings, we could for example choose to consider the net power entering and exiting the particle, i.e.,
 \begin{align}
 \hat{f}(\tau)
   =
   \hat{\xi}(\tau)
   \delta(t-\tau)
   -
   \gamma(t-\tau)\dot{\hat{Q}}(\tau),
 \end{align}
 where the first term corresponds to the incoming power and the second term to the outgoing power.
 For simplicity, we focus on the contributions to the power connected to the noise operator ($P_{\mathrm{in}}$ and $P_{\mathrm{out}}^{\mathrm{fluc}}$), as the contributions connected to the initial conditions simply decay into the environment over time (see Equation \eqref{Eq:PowerDef2} and discussion below).
 Given the specific form of $\hat{f}$, due to the self-consistency of our system, any autocorrelation of $\hat{J}$ reduces to calculating the moments of the (Gaussian) noise operator $\hat{\xi}$ and solving the subsequent time integral [see Equation \eqref{Eq:HigherOrderCorr}].
 For the net power the system exchanges with the environment, we obtain
 \begin{align}
 \label{Eq:JTotalPower}
 \hat{J}
   &=
   \hat{J}_{\mathrm{in}}
   -
   \hat{J}_{\mathrm{out}}^{\mathrm{fluc}}
   \\\nonumber
   &
   =
   \frac{1}{2}
   \int_0^t\mathrm{d}s~
 	\\\nonumber
 	&\times
   \left[
 	\{
   \hat{\xi}(s)
 	,\dot{\hat{\mathcal{Q}}}(s)
 	\}
   -
   2\int_0^s\mathrm{d}\tau~\gamma(s-\tau)
 	\{\dot{\hat{\mathcal{Q}}}(\tau)
 	,
 	\dot{\hat{\mathcal{Q}}}(s)
 	\}
   \right].
 \end{align}
 We remark that the clear splitting in different contributions of energy flows is connected to the linearity of the system. By using the symmetric average, we can render the individual parts as physically meaningful. This has been extensively discussed in the literature, and we refer the reader to refs. \cite{dalibard82,li93,senitzky95,li95} for further details.
 In Figure \ref{fig:mesh1}, we report the outgoing power connected to the system's fluctuating dynamics using the bath spectral density of Equation \eqref{Eq:ILambda} (second part of Equations \eqref{Eq:JTotalPower}, see also Equation \eqref{Eq:PoutTimeDomain}).
 At late times, where the system can equilibrate, it balances the ingoing power (first part of Equation~\eqref{Eq:JTotalPower}, see also Equation \eqref{Eq:PowerDef}), and there is no net transfer of energy between system and its environment on average \cite{hsiang15,barton16}.
 During the full course of the nonequilibrium evolution, the corresponding expression using the thermal fluctuation--dissipation inequality [Equation~\eqref{Eq:ThermalInequality}] provides a lower bound to the outgoing power.
 Since the properties of the momentum correlation are inherited by the power, this behavior can be understood in full similarity to our discussion of the momentum correlations (see Equation \eqref{Eq:DeltaqDot} and discussion below).

 \subsection{Generalized Current Fluctuations}

 In the context of the thermodynamic uncertainty relation, it is more interesting for us to consider the \emph{{fluctuations}
 } of the generalized current {(see Section \ref{Sec:Introduction})}.
 As an illustration and to keep the expressions transparent, we focus on the incoming power only and consider the total input power, i.e.,
 \begin{align}
 \label{Eq:IncomingP}
 \hat{f}
   &=\hat{\xi}(\tau)
   \rightarrow
   \hat{J}_{\rm in}
   =
   \frac{1}{2}
   \int_0^t
   \mathrm{d}\tau~
   \{
   \hat{\xi}(\tau)
   ,
   \dot{\hat{Q}}(\tau)
   \}.
 \end{align}
 On average, as the integral kernel of the previous equation approaches a constant at late times (see Section \ref{Sec:EnergyFlow}), the mean current is equal to the incoming instantaneous power, i.e., $\langle\hat{J}\rangle_{\rm s}/t\to P_{\rm in}$ for $t\to\infty$ {(we use the time average for convergence following Refs. \cite{saito11,saryal21})}.
 Indeed, using that
 $
 \int_0^t\mathrm{d}\tau\int_0^{\tau}\mathrm{d}x=\int_0^t\mathrm{d}x\int_x^t\mathrm{d}\tau
 $, we have that
 \begin{align}
 \frac{\langle\hat{J}_{\rm in}\rangle_{\rm s}}{t}
   &=
   \int_0^t
   \frac{\mathrm{d}\tau}{t}
   \int_0^{\tau}
   \mathrm{d}x~
   \dot{\chi}(x)
   \nu(x)
   \\\nonumber
   &=
   \int_0^{t}
   \mathrm{d}x~
   \frac{t-x}{t}
   \dot{\chi}(x)
   \nu(x)
   \to
   P_{\rm in}(t), \qquad t\to\infty.
 \end{align}
 For finite times, on the other hand, using Equation \eqref{Eq:HigherOrderCorr}, we obtain
 \begin{align}
 \label{Eq:CurrentFluctuations2}
 &\frac{\langle (\Delta\hat{J}_{\rm in})^2\rangle_{\rm s}}{t}
   =
   \int_0^t\frac{\mathrm{d}x}{t}
   \int_0^t\mathrm{d}y~
   \\\nonumber
   &
   \times
   \left[
   \langle\hat{\xi}(x)\dot{\hat{Q}}(y)\rangle
   \langle\dot{\hat{Q}}(x)\hat{\xi}(y)\rangle
   +
   \langle\hat{\xi}(x)\hat{\xi}(y)\rangle
   \langle\dot{\hat{Q}}(x)\dot{\hat{Q}}(y)\rangle
   \right].
 \end{align}
 This result is exact for all times of the nonequilibrium evolution and we note that we do not take the symmetric average on the right-hand side of the previous line.
 In order to employ our results on the (thermal) fluctuation--dissipation inequality, we need to reorder Equation~\eqref{Eq:CurrentFluctuations2} using the commutators.
 \begin{subequations}
 \label{Eq:Commutator}
 \begin{align}
 \langle
 [\hat{\xi}(x),\dot{\hat{Q}}(y)]
 \rangle
   &=
   2i\hbar
   \int_0^y\mathrm{d}\tau~
   \dot{\chi}(y-\tau)
   \Gamma(x-\tau),
   \\
   \langle
   [\dot{\hat{Q}}(x),\dot{\hat{Q}}(y)]
   \rangle
     &=
     2i\hbar
     \int_0^x\mathrm{d}\tau
     \int_0^y\mathrm{d}\bar{\tau}~
     \\\nonumber
     &\qquad\times
 		\dot{\chi}(x-\tau)
     \dot{\chi}(y-\bar{\tau})
     \Gamma(\tau-\bar{\tau})
   ,
 \end{align}
 \end{subequations}
 which yields for the variance of the generalized current operator
 \begin{align}
 \label{Eq:CurrentFluctuations3}
 &\frac{\langle (\Delta\hat{J}_{\rm in})^2\rangle_{\rm s}}{t}
   =
   \int_0^t
   \frac{\mathrm{d}x}{t}
   \int_0^t
   \mathrm{d}y~
   \\\nonumber
   &\times
   \left[
   \langle
   \hat{\xi}(x)
   \dot{\hat{Q}}(y)
   \rangle_{\rm s}
   \left(
   \langle
   \dot{\hat{Q}}(x)
   \hat{\xi}(y)
   \rangle_{\rm s}
   +
   \frac{1}{2}
   \langle
   [\dot{\hat{Q}}(x),
   \hat{\xi}(y)]
   \rangle
   \right)
   \right.
   \\\nonumber
   &\quad +
   \nu(x-y)
   \left(
   \langle
   \dot{\hat{Q}}(x)
   \dot{\hat{Q}}(y)
   \rangle_{\rm s}
   +
   \frac{1}{2}
   \langle
   [\dot{\hat{Q}}(x),
   \dot{\hat{Q}}(y)]
   \rangle_{\rm s}
   \right)
   \\\nonumber
   &
   \quad+
   \frac{1}{2}
   \langle
   [
   \hat{\xi}(x)
   ,
   \dot{\hat{Q}}(y)
   ]
   \rangle
   \left(
   \langle
   \dot{\hat{Q}}(x)
   \hat{\xi}(y)
   \rangle_{\rm s}
   +
   \frac{1}{2}
   \langle
   [
   \dot{\hat{Q}}(x)
   ,
   \hat{\xi}(y)
   ]
   \rangle
   \right)
   \\\nonumber
   &\left.\quad
   +
   \frac{1}{2}
   \langle
   [
   \hat{\xi}(x)
   ,
   \hat{\xi}(y)
   ]
   \rangle
   \left(
   \langle
   \dot{\hat{Q}}(x)
   \dot{\hat{Q}}(y)
   \rangle_{\rm s}
   +
   \frac{1}{2}
   \langle
   [
   \dot{\hat{Q}}(x)
   ,
   \dot{\hat{Q}}(y)
   ]
   \rangle
   \right)
   \right].
 \end{align}
 The previous line might appear more complicated than Equation~\eqref{Eq:CurrentFluctuations2}, but it enables us to dissect the underlying physics.

 Firstly, the foremost term of Equation \eqref{Eq:CurrentFluctuations3} is related to the incoming power $P_{\rm in}(t)$ in the frequency domain.
 Writing
 $
 P_{\rm in}(t)
   =
   \int_0^t\mathrm{d}\tau
   \int\frac{\mathrm{d}\omega}{2\pi}
   \int\frac{\mathrm{d}\bar{\omega}}{2\pi}
   (-i\omega)\alpha(\omega)
   \nu(\bar{\omega})
   e^{-i[\omega+\bar{\omega}]\tau}
 $, we can find a relation to the \emph{frequency} components of the incoming power, i.e.,
 \begin{align}
 &\langle
 \hat{\xi}(x)
 \dot{\hat{Q}}(y)
 \rangle_{\rm s}
   =
   \int_0^y\mathrm{d}\tau~
   \dot{\chi}(y-\tau)
   \nu(x-\tau)
   \\\nonumber
   &=
   \int_0^y\mathrm{d}z
   \int\frac{\mathrm{d}\omega}{2\pi}
   \int\frac{\mathrm{d}\bar{\omega}}{2\pi}
   (-i\omega)\alpha(\omega)
   \nu(\bar{\omega})
   e^{-i(\omega+\bar{\omega}) z}
   e^{-i\bar{\omega}(x-y)}.
 \end{align}
 In other words, the first part of Equation \eqref{Eq:CurrentFluctuations3} can be connected to the (squared) incoming power via the integral kernel
 $
 	(-i\omega)\alpha(\omega)\nu(\bar{\omega})
 $.
 This becomes clear at late times, where we can write
 \begin{align}
 \lim_{t\to\infty}
 \int_0^t
 &\frac{\mathrm{d}x}{t}
 \int_0^t
 \mathrm{d}y~
 \langle
 \hat{\xi}(x)
 \dot{\hat{Q}}(y)
 \rangle_{\rm s}
 \langle
 \hat{\xi}(y)
 \dot{\hat{Q}}(x)
 \rangle_{\rm s}
   \\\nonumber
   &\sim
   \lim_{t\to\infty}
   \int_{0}^t
   \frac{\mathrm{d}x}{t}
   \int\frac{\mathrm{d}\omega}{2\pi}
   \left[
   (-i\omega)
   \alpha(\omega)
   \nu(\omega)
   \right]^2
   \\\nonumber
   &=
   2
   \int_0^{\infty}\frac{\mathrm{d}\omega}{2\pi}
   \text{Re}[
   \{(-i\omega)
   \alpha(\omega)
   \nu(\omega)
   \}^2],
 \end{align}
 where we
 used that $\nu(\omega)$ is an even function and that $\alpha(\omega)$ fulfills the Kramers--Kronig relations \cite{jackson99}.
 For comparison, the incoming power at late times can be written as
 $
 \lim_{t\to\infty}P_{\rm in}(t)
   =2
   \int_0^{\infty}\mathrm{d}\omega~P_{\rm in}(\omega),
 $
 where we defined
 $
 P_{\rm in}(\omega)
   =
   \text{Re}[(-i\omega)
   \alpha(\omega)
   \nu(\omega)]
 $.
 We further note that we can explicitly see that the incoming power balances the outgoing power at late times.
 To this end, we write
 $
 \alpha_I
   =
   2\omega\text{Re}[\gamma_{\theta}(\omega)]|\alpha(\omega)|^2
 $.
 For the outgoing power, on the other hand, we obtain
 $
 P_{\mathrm{out}}
   \to
   2\int_0^{\infty}\mathrm{d}\omega~
   P_{\mathrm{out}}(\omega)
   =P_{\mathrm{in}}
 $
 with $
 P_{\mathrm{out}}(\omega)
   =
   2\omega^2\text{Re}[\gamma_{\theta}(\omega)]|\alpha(\omega)|^2\nu(\omega)
   =
   P_{\rm in}(\omega)
   $
 in accordance with detailed balance.

 Secondly, the last two lines of Equation~\eqref{Eq:CurrentFluctuations3} feature two terms that are given by average commutators only (last term in brackets in the last two lines).
 These are temperature-independent (see Equation \eqref{Eq:Commutator} in combination with Equation \eqref{Eq:FDI} and discussion below) and are hence solely dependent on quantum fluctuations.
 In the high-temperature (classical) limit, where thermal fluctuations dominate, they can be ignored.

 Thirdly, Equation~\eqref{Eq:CurrentFluctuations3} contains a number of terms that are given by a combination of a symmetric average and an average of a commutator.
 For times larger than the typical dissipation time scale of the system ($\sim$$\gamma$), these can be expected to become exponentially small due to symmetry reasons under the integral.

 Lastly, the remaining term (first term of the second line) of Equation~\eqref{Eq:CurrentFluctuations3} is even more interesting from our perspective, as the noise kernel is evaluated over the two-time correlations of the momentum operator, and we can determine a lower bound by means of the thermal fluctuation--dissipation inequality [Equation~\eqref{Eq:ThermalInequality}].
 To this end, we split the integral into
 $
 \int_0^t
 \mathrm{d}x
 \int_0^t
 \mathrm{d}y
   =
   \int_0^t
   \mathrm{d}x
   \left[
   \int_0^x
   \mathrm{d}y
   +
   \int_x^t
   \mathrm{d}y
   \right]
 $
 and transform
 $
 \int_0^t
 \mathrm{d}x
 \int_x^t
 \mathrm{d}y
   =
   \int_0^t
   \mathrm{d}y
   \int_0^y
   \mathrm{d}x
 $.
 The first term of the second line in Equation \eqref{Eq:CurrentFluctuations3} can then be written as
 \begin{align}
   \int_0^t
   \frac{\mathrm{d}x}{t}
   &
   \int_0^t
   \mathrm{d}y~
   \nu(x-y)
   \langle
   \dot{\hat{Q}}(x)
   \dot{\hat{Q}}(y)
   \rangle_{\rm s}
   \\\nonumber
   &
   =
   2
   \int_0^t
   \frac{\mathrm{d}x}{t}
   \int_0^x
   \mathrm{d}y~
   \nu(x-y)
   \langle
   \dot{\hat{Q}}(x)
   \dot{\hat{Q}}(y)
   \rangle_{\rm s}
   ,
 \end{align}
 where we used that the integral kernel is invariant with respect to replacing $x\leftrightarrow y$.
 Again, at later times, where the initial jolt has been damped and the system is approaching the steady-state, we can employ the thermal fluctuation--dissipation inequality (Equation \eqref{Eq:ThermalInequality}) and recall the expression for the outgoing power
 $
 P_{\mathrm{out}}^{\rm fluc}(t)
 	=
   2
 	\int_0^t\mathrm{d}\tau~
 	\gamma(t-\tau)
   \langle
 	\dot{\hat{\mathcal{Q}}}(\tau)
 	\dot{\hat{\mathcal{Q}}}(t)
   \rangle_{\mathrm{s}}
 $ [see Equation \eqref{Eq:PoutTimeDomain}],
 in order to write
 \begin{align}
 \label{Eq:CurrentInequality3}
 2
 \int_0^t
 \frac{\mathrm{d}x}{t}
 \int_0^x
 \mathrm{d}y~
 &
 \nu(x-y)
 \langle
 \dot{\hat{Q}}(x)
 \dot{\hat{Q}}(y)
 \rangle_{\rm s}
 \\\nonumber
   &
 \geq
   \frac{2}{\beta}
   \int_0^t
   \frac{\mathrm{d}\tau}{t}~
   P_{\rm out}^{\rm fluc}(\tau)
   \\\nonumber
   &
   =
   \frac{2}{\beta}
   \frac{
   \langle \hat{J}_{\rm out}^{\rm fluc}\rangle_{\rm s}}{t}
   .
 \end{align}

 Collecting the results from the previous points, we can formulate the relation
 \begin{align}
 \label{Eq:TURApprox}
 \frac{\langle (\Delta\hat{J}_{\rm in})^2\rangle_{\rm s}}{\langle \hat{J}_{\rm out}^{\rm fluc}\rangle_{\rm s}^2}
 \langle \hat{J}_{\rm out}^{\rm fluc}\rangle_{\rm s}
   \gtrsim
   2k_{\rm B}T
 \end{align}
 which is a \emph{{new}} form of thermodynamic uncertainty relations: the fluctuations of the generalized current are greater than or equal to the (time-averaged) outgoing power times a thermal factor of $2k_{\rm B}T$. From the perspective of the Brownian particle, the latter can be interpreted as dissipation/loss of energy to the environment. Increasing the accuracy of the incoming power comes at least at the price of the time-averaged dissipated energy (compare with the original formulation of the TUR in ref. \cite{barato15}).
 Establishing an unambiguous connection to entropy production in the complete system---although rather commonly invoked in conventional formulations of TURs---is nontrivial for general non-Markovian quantum systems under fully nonequilibrium conditions.
 We refer the reader, e.g., to our first work \cite{dong22} and references therein for more details. Instead, we  outline in Section \ref{Sec:HeatTrans} how our formalism can be applied to steady-state heat transfer where an unambiguous connection to entropy production is possible. To derive Equation \eqref{Eq:TURApprox}, we had to assume that (i) the transitional dynamics is starting to settle, i.e., we work at times larger than the characteristic damping scales of the system+environment composite approaching the steady state (equilibrium in the current case), and (ii) that quantum fluctuations can be ignored.
 In any other situation, one needs to exercise particular care and is probably better advised to use Equation \eqref{Eq:CurrentFluctuations3}.

 \subsection{Non-Markovianity of the Damping Kernel}

 Remarkably, our thermodynamic uncertainty relation connects the fluctuations of the nonequilibrium current connected to the incoming power operator $\hat{J}_{\rm in}$ to the power leaving the system $P_{\rm out}^{\rm fluc}$.
 It thereby establishes a statistical statement on the energetic interaction between system and environment in the course of their equilibration process.
 Only at late times, where $P_{\rm out}\to P_{\rm in}$, the incoming power itself can be related to the lower thermodynamic bound of its corresponding fluctuations.
 Since we could provide an exact relation [Equation \eqref{Eq:CurrentFluctuations3}], the inequality can be refined by (i) including the additional terms from Equation \eqref{Eq:CurrentFluctuations3} or (ii) by including higher-order corrections from the noise correlation (Equation \eqref{Eq:NuIntegral}).

 Furthermore, we would like to comment on the Markovian and high-temperature limit of Equation \eqref{Eq:CurrentInequality3}. Starting from the $\Lambda$-model for the bath spectral density (Equation~\eqref{Eq:LambdaGamma} and Appendix \ref{App:BathSpectrDens}), the Markovian case is achieved in the limit $\Lambda\to\infty$, where $\gamma_{\Lambda}(\tau)\to2\gamma_0\delta(\tau)$.
 In this case, the fluctuating part of the outgoing power reduces to
 $
 P_{\mathrm{out}}^{\rm fluc}(t)
 	\to
   2
 	\gamma_0
   \langle
 	\dot{\hat{\mathcal{Q}}}^2(t)
   \rangle_{\mathrm{s}}
 $ and is fully determined by the momentum fluctuations of the system.
 For comparison, we have seen earlier (Figure \ref{fig:qdot}) that it is generally true that
 $
 \langle
 \dot{\hat{\mathcal{Q}}}^2(t)
 \rangle_{\mathrm{s}}\geq
 \langle
 \dot{\hat{\mathcal{Q}}}(t)
 \dot{\hat{\mathcal{Q}}}(\tau)
 \rangle_{\mathrm{s}}
 $, i.e., the momentum correlations become maximal at equal times.
 Hence, a general upper bound for the incoming power is given by
 \begin{align}
 \label{Eq:PowerOutUpperLimit}
 P_{\rm out}^{\rm fluc}
   \leq
   2
   \langle
   \dot{\hat{\mathcal{Q}}}^2(t)
   \rangle_{\mathrm{s}}
   \int_0^t\mathrm{d}\tau ~\gamma(t-\tau)
   .
 \end{align}
 Choosing once again $\gamma=\gamma_{\Lambda}$ in the limit $\Lambda\to\infty$, we would reproduce the Markovian result for $P_{\rm out}^{\rm fluc}$. However, for the non-Markovian case with finite $\Lambda$, we instead obtain
 $
 P_{\rm out}^{\rm fluc}
   \leq
   2
   \gamma_0
   (1-e^{-\Lambda t})
   \langle
   \dot{\hat{\mathcal{Q}}}^2(t)
   \rangle_{\mathrm{s}}
 $
 ($t\geq 0$).
 This is less than or equal to the Markovian limit, and the equality is only achieved at late times ($t\to\infty$) (see Figure \ref{fig:mesh1}).
 In other words, our non-Markovian thermodynamic uncertainty relation (Equation \eqref{Eq:CurrentInequality3}) can be expected to provide a lower bound than its Markovian counterparts.
 Furthermore, we remark that Equation~\eqref{Eq:PowerOutUpperLimit} is sensitive to the particular regularization scheme \cite{hsiang18,dong22}.

 Lastly, let us once again emphasize that it is not the particular example of the nonequilibrium current we are interested in, but Equation \eqref{Eq:IncomingP} was rather chosen for its simple form.
 The important insight is that we could deduce a relation from a chain of arguments starting from the hermicity of the noise operators and the self-consistency of our system+bath dynamics over the uncertainty relations and the fluctuation--dissipation inequality, all the way to what we coined the thermal thermodynamic uncertainty relation. In this way, the thermodynamic uncertainty relation finds its clear footing in the microscopic stochastic properties of open quantum systems.

 We conclude our discussion in the next section by discussing one more example and connecting our formalism to the related problem of steady-state heat transfer.

 \section{Nonequilibrium Steady State and Connection to Heat Transfer}\label{Sec:HeatTrans}

 At late times, the energy lost to the environment and the energy inflow into the system balance on average such that the system can equilibrate \cite{hsiang18}.
 The net current (Equation~\eqref{Eq:JTotalPower}) vanishes in the mean, i.e.,
 $
 \langle\hat{J}\rangle_{\rm s}=0
 $.
 This, however, is not true for its fluctuations that can remain finite even in equilibrium.
 To see this more clearly, let us again for simplicity consider the current connected to the incoming power only (Equation~\eqref{Eq:IncomingP}).
 From Equation~\eqref{Eq:CurrentInequality3}, we immediately find the expected late-time limit
 $
 (2/\beta)\langle\hat{J}_{\rm out}^{\rm fluc}\rangle_{\rm s}
   \to (2/\beta)\lim_{t\to\infty}P_{\rm out}(t)
 $
 as found in Equation \eqref{Eq:POutFluc}
 if we use the relation
 $
 \lim_{t\to\infty}
 \int_0^t
 \frac{\mathrm{d}x}{t}
 \int_0^x\mathrm{d}y~
 e^{-iy\omega}
   =
   \lim_{\epsilon\to 0}
   \frac{-i}{\omega-i\epsilon}
 $.

 The formalism we used above can be readily extended to a situation of multiple particles connected to multiple heat baths at different temperatures \cite{dhar03,hsiang15,barton16}.
 At late times, the system then does  not equilibrate but rather reaches a nonequilibrium steady state, where it mediates a constant average heat transfer between the heat baths \cite{rieder67,polder71} (see also refs. \cite{hsiang15,barton16,lopez18,pekola21} for a modern take on the topic).
 The fluctuations of such a nonequilibrium steady-state current have been calculated quite generally in refs. \cite{saito07,zhan11,dechant18}, and a particular focus on the thermodynamic uncertainty relation was put just recently in refs. \cite{saryal19,saryal21}.
 Comparing with our full nonequilibrium form in Equation \eqref{Eq:CurrentFluctuations2}, it can be instructive to explore how the thermal fluctuation--dissipation inequality affects the thermodynamic uncertainty in steady-state heat transfer.
 It again turns out to be the hidden fundamental principle determining the macroscopic statistical behavior.

 For simplicity, we consider two interacting particles connected to two heat baths individually.
 Furthermore, we consider the simplified example of a delta-function-shaped dissipation kernel
 $
 \gamma(t)\sim\gamma_0\delta(t).
 $
 Following the approach and notation of ref. \cite{hsiang15}, we then obtain a system of two coupled quantum Langevin equations (both particles feature equal mass $m=1$), i.e.,
 \begin{align}
 \label{Eq:LangevinCoupledOsc}
 \ddot{\hat{\bm{\chi}}}
   +2\gamma_0
   \dot{\hat{\bm{\chi}}}
   +
   \underline{\Omega}\hat{\bm{\chi}}
   =
   \hat{\bm{\xi}},
 \end{align}
 where
 $\underline{\Omega}=
 \left(\begin{smallmatrix}\omega_0^2 & \sigma\\\sigma & \omega_0^2\end{smallmatrix}\right)$ is the coupling matrix, $\hat{\bm{\chi}}=(\hat{\chi}_1,\hat{\chi}_2)^T$ the vector operator for the two quantum degrees of freedom, and
 $\hat{\bm{\xi}}=(\hat{\xi}_1,\hat{\xi}_2)^T$ the corresponding noise operator of the respective quantum baths with
 $
 \langle
 \hat{\xi}_i(\omega)
 \hat{\xi}_j(\omega')
 \rangle_{\rm s}
   \equiv
   2\pi\delta(\omega+\omega')
   [\underline{G}_H(\omega)]_{ij}
   =
   4\pi\delta(\omega+\omega')
   \delta_{ij}
   \gamma_0
   \hbar\omega\coth[\hbar\omega\beta_i/2]
 $.
 Here, $\delta_{ij}$ is the Kronecker-delta and {$\beta_{1|2}$} the inverse temperature of the respective quantum baths.
 The heat flow can be measured at various different points of the system.
 However, in the nonequilibrium steady state, the particular choice does not influence the result anymore.
 For convenience, we define the nonequilibrium current as the energy transferred from particle 2 to particle 1, i.e.,
 \begin{align}
 \label{Eq:CurrentChoice}
 \hat{J}
   &=
   -
   \frac{\sigma}{2}
   \int_0^t\mathrm{d}s~
   \{\hat{\chi}_2(s),\dot{\hat{\chi}}_1(s)\}.
 \end{align}
 We note that earlier references, e.g., refs. \cite{saito07,pagel13,saryal19}, considered the energy transferred from one reservoir into the coupled system, which requires a somewhat different calculation.
 In Appendix \ref{App:Heattrans}, we explicitly show that the same result can be obtained with the point of measurement in Equation \eqref{Eq:CurrentChoice}.
 After a lengthy but straight-forward calculation, we find
 \begin{align}
 \label{Eq:FlucHeatTransmission}
 \frac{\langle(\Delta\hat{J})^2\rangle_{\rm s}}{t}
     &\sim
     2
     \int\frac{\mathrm{d}\omega}{2\pi}~
     (\hbar\omega)^2T(\omega)
     \\\nonumber
     &\times
     \left[
     T(\omega)
     \left(
     \coth\left[\frac{\hbar\omega\beta_1}{2}\right]
     -
     \coth\left[\frac{\hbar\omega\beta_2}{2}\right]
     \right)^2
     \right.
     \\\nonumber
     &\quad+
     \left.
     \frac{1}{2}
     \coth\left[\frac{\hbar\omega\beta_1}{2}\right]
     \coth\left[\frac{\hbar\omega\beta_2}{2}\right]
     -\frac{1}{2}
     \right]
 \end{align}
with the transmission function
 $T(\omega)
   =
   (\gamma_0\omega\sigma)^2
   \Pi_{\pm}
   [\omega_0^2-\omega^2\pm\sigma\pm 2i\gamma_0\omega]^{-1}$ (the product runs over all possible combination of $\pm1$)
 This result, in different contexts leading to various transmission coefficients, has been found by many authors before, see, e.g., \cite{saito07,saito11,zhan11,pagel13,tang18} and references therein.

Recently, it was found \cite{saryal19,saryal21} that  steady-state heat transfer is in accordance with the thermodynamic uncertainty relation by means of an expansion of the thermal functions in Equation \eqref{Eq:FlucHeatTransmission}, independent of the concrete form of the transmission coefficient (see Equations (13)--(15) in \cite{saryal19}).
 To this end, we use that
 $
 \coth[x/2]=1+2N(x)
 $
 with $N(x)=[e^x-1]^{-1}$ as the bosonic occupation number.
 If we further neglect the strictly positive first term in Equation \eqref{Eq:FlucHeatTransmission}, we follow ref. \cite{saryal21} and readily find
 \begin{align}
 \label{Eq:FlucHeatTransmission2}
 \frac{\langle(\Delta\hat{J})^2\rangle_{\rm s}}{t}
     &\sim
     2
     \int\frac{\mathrm{d}\omega}{2\pi}~
     (\hbar\omega)^2T(\omega)
     \\\nonumber
     &\times
     \coth\left[\frac{\hbar\omega}{2(\beta_2-\beta_1)}\right]
     [N(\beta_1)-N(\beta_2)]
     \\\nonumber
     &
     \geq
     \frac{2}{\beta_2-\beta_1}
     \frac{\langle\hat{J}\rangle_{\rm s}}{t}.
 \end{align}
 Although the previous result could have been similarly obtained from purely mathematical arguments, it becomes clear from our present discussion that the thermodynamic uncertainty relation in steady-state heat transfer is deeply rooted in the fluctuation--dissipation inequality, and hence the causality and the statistical properties of the underlying Hamiltonian.
 Indeed, in our simple case of heat transfer with time-local dissipation kernel, the thermal fluctuation--dissipation inequality is practically a statement on the properties of the hyperbolic cotangent in the second line of Equation \eqref{Eq:FlucHeatTransmission} (see discussion below Equation~\eqref{Eq:ThermalInequality}), which was the necessary step in order to derive the inequality in Equation~\eqref{Eq:FlucHeatTransmission2}.

 Lastly, we comment on the connection between the thermodynamic uncertainty relation and entropy production in the system. In the nonequilibrium steady state, the density matrix of the system of interest becomes stationary per definition, as does the von Neumann entropy of the system.
 However, due to the constant energy flow mediated by the system, the total entropy of the system+bath increases at a constant rate \cite{colla21,talkner20} and the average entropy production rate in the total system was identified via $\langle\partial_t\mathcal{S}\rangle_{\rm s}=(\beta_2-\beta_1)^{-1}\langle\hat{J}\rangle_{\rm s}$ \cite{saryal19,saryal21}.
 In the case of quantum Brownian motion, where the system can equilibrate at late times, the entropy production vanishes.
 It is rather the total change in entropy {of the system} over the course of the equilibration process (Equation \eqref{Eq:Entropy}) that quantifies {the uncertainties}. From the perspective of the thermodynamic uncertainty relation, it can then be more convenient to use the power flow between system and environment in order to quantify dissipation (Equation~\eqref{Eq:CurrentInequality3}).

 \section{Conclusions}\label{Sec:Summary}

 We systematically explored the emergence of macroscopic thermodynamic uncertainty relations \cite{barato15} for generalized currents from microscopic uncertainty principles of interacting open quantum systems.
 Using the generic example of quantum Brownian motion, we set the microscopic framework by adopting the fluctuation--dissipation inequality \cite{fleming13} for the system's microscopic quantum degrees of freedom (canonical operators) \cite{dong22}.
 The fluctuation--dissipation inequality is solely based on reasonable physical assumptions, such as the hermicity of the involved operators and a causal interaction between system and bath.
 It provides the inexpugnable lower bound for the system's microscopic uncertainty---including the Robertson--Schr\"odinger inequality as the nonequilibrium standard quantum limit.
 However, when thermal fluctuations are dominating, the fluctuation--dissipation inequality may not provide the most accurate lower bound.
 Moreover, although the fluctuation--dissipation inequality is a fairly fundamental statement, it is not sufficient to fully explain the emergence of a thermodynamic uncertainty relation for macroscopic thermodynamic quantities.
 Instead, we needed to specify the statistical properties of the bath in order to extract information on the temperature dependence of the system's fluctuations.
 Based on a microscopic model for the bath spectral density, this led us to formulating a \emph{{thermal}
 } fluctuation--dissipation relation that is valid at high temperatures, which provides a tighter bound than the fluctuation--dissipation inequality.
 This means that it can provide a better estimate {in the thermal-energy-dominated regimes} (see Equation~\eqref{Eq:ThermalInequality}).
 This thermal fluctuation--dissipation inequality {which applies to the classical domain enables us to compare with popular TURs in the literature \cite{barato15,vu19,horowitz20,dechant21}, where the high-temperature limit is a requirement rather than a particular limit}.
 When applied to the generalized current, we show how it formally leads to a thermodynamic uncertainty relation.
 In essence, it is the combination of the statistical properties of the bath and the causality of the system+bath interaction that is inherited by thermodynamic quantities (e.g. generalized currents) and can be seen as the microscopic origin of thermodynamic uncertainty relations in linear open quantum systems.

 As an instructive example, we examine the power entering the Brownian particle as the generalized current and find an exact expression of the corresponding current--current correlations.
 At high temperatures, applying the thermal fluctuation--dissipation inequality, {in Equation \eqref{Eq:TURApprox} we showed} that the corresponding fluctuations are always equal to or greater than the average \emph{{outgoing}
 } power weighted by a thermal factor $2k_{\rm B}T$.
 For temperatures that are low with respect to the dominating energy scales in the system, quantum corrections need to be taken into account.
 Our result fully includes back-action from the environment onto the system and respects the particular anatomy of realistic bath spectral densities, albeit limited to the linear regime.
 It is straight-forward to extend our formalism to multiple particles connected to several heat baths.
 This enabled us to check the consistency of our result by rederiving a thermodynamic uncertainty relation already known from the literature on steady-state heat transfer.

Our analysis sheds some light on how thermodynamic uncertainty relations are deeply rooted in the quantum uncertainty relations and how they impact  macroscopic observables.
It further provides some basic tests and requirements in the form of inequalities that any trustworthy physical result within the realm of our assumptions must comply with.

 ~~\textit{Acknowledgements}~
 D. Reiche thanks Francesco Intravaia, Kurt Busch, and Markus Krutzik for enlightening discussions.
 J.-T. Hsiang and B.-L. Hu thank Hing-Tong Cho for an independent check of results presented in Appendix \ref{App:WicksTheorem}.
 DR gratefully acknowledges support from the German Space Agency (DLR) with funds provided by the Federal Ministry for Economic Affairs and Climate Action (BMWK) under grant number 50MW2251.
 JTH is supported by the Ministry of Science and Technology of Taiwan, R.O.C., under Grant No. MOST 110-2811-M-008-522.
 The article processing charge was funded by the Deutsche Forschungsgemeinschaft (DFG, German Research Foundation)--491192747 and the Open Access Publication Fund of Humboldt-Universität zu Berlin.


 \appendix

 \section{Langevin Equation and Uncertainty Relations}\label{SolvLangevin}

 In the following, we provide details on the properties of the Langevin equation and its solutions discussed in Section \ref{Sec:StochasticDynamics}.

 The response function $\chi(t)$ obeys the homogeneous equation
 \begin{align}
 \label{Eq:GreenHomII}
 \ddot{\chi}(t)
 	+
 	\omega_0^2\chi(t)
 	+2
 	\int_{-\infty}^t\mathrm{d}\tau~
 	\gamma_{\theta}(t-\tau)\dot{\chi}(\tau)=0,
 \end{align}
 with the initial values $\chi(0)=0$, $\dot{\chi}(0)=1$.
 We note that the response function is causal $\chi(\tau)=0 \text{ for } \tau<0$.
 Furthermore, we can transform the convolution in Equation \eqref{Eq:GreenHomII} to an algebraic expression in Fourier domain and obtain for the response function the expression in Equation \eqref{Eq:GreenFourier}.
 Defining the matrix
 \begin{align}
 \label{Eq:DefX}
 \underline{X}(t)
 	=
 	\begin{pmatrix}
 	\dot{\chi}(t) & \chi(t)\\
 	\ddot{\chi}(t) & \dot{\chi}(t)
 	\end{pmatrix},
 \end{align}
 The general time-dependent solution for the quantum degree of freedom reads
 \begin{align}
 \label{Eq:qSol}
 \mathbf{Q}(t):=
 \binom{\hat{q}(t)}{\dot{\hat{q}}(t)}
 	&=
 	\underline{X}(t)
 	\binom{\hat{q}(0)}{\dot{\hat{q}}(0)}
 	+
 	\int_0^t\mathrm{d}\tau~
 	\underline{X}(t-\tau)
 	\binom{0}{\hat{\xi}(\tau)}.
 \end{align}
 From this, we can derive the second-order correlations in Equation \eqref{Eq:FluctSystem}.

 \subsection{Fluctuation--Dissipation Inequality}\label{Sec:FDI}

 We begin our discussion with the relation between $\nu$ and $\gamma$ characterizing the environment.
 For any quantum mechanically well-defined operator $\hat{O}(t)$ and scalar product $\langle\cdot\rangle$, we can state that
 \begin{align}
 \langle\hat{O}^{\dagger}(t)\hat{O}(t)\rangle\geq 0.
 \end{align}
 This implies that the quantum average of $\hat{O}^{\dagger}(t)\hat{O}(t)$ is a semipositive definite function of time in the sense of~\cite{ford88,fleming13}.
 \begin{align}\label{Eq:DefPosDefinite}
 \int_0^t&\mathrm{d}\tau
 \int_0^t\mathrm{d}\tau'~
 f^*(\tau)
 \langle\hat{O}^{\dagger}(\tau)\hat{O}(\tau')\rangle
 f(\tau')
 \\\nonumber
 	&
 =
 	\left\langle
 	\left(
 	\int_0^t\mathrm{d}\tau
 	f(\tau)
 	\hat{O}(\tau)
 	\right)^{\dagger}
 	\int_0^t\mathrm{d}\tau'~
 	f(\tau')
 	\hat{O}(\tau')
 	\right\rangle
 	\geq 0
 \end{align}
 for all complex functions $f(t)$.
 Writing
 $
 \hat{O}^{\dagger}(t)\hat{O}(t)
 	=
 	(
 	\{\hat{O}^{\dagger}(t),\hat{O}(t)\}
 	+
 	[\hat{O}^{\dagger}(t),\hat{O}(t)]
 	)/2
 $,
 we have that
 \begin{align}
 \label{Eq:FDI2}
 \int_0^t&\mathrm{d}\tau
 \int_0^t\mathrm{d}\tau'~
 f^*(\tau)
 \langle\{\hat{O}^{\dagger}(\tau),\hat{O}(\tau')\}\rangle
 f(\tau')
 \\\nonumber
 	&
 \geq
 	\pm
 	\int_0^t\mathrm{d}\tau
 	\int_0^t\mathrm{d}\tau'~
 	f^*(\tau)
 	\langle[\hat{O}^{\dagger}(\tau),\hat{O}(\tau')]\rangle
 	f(\tau').
 \end{align}
 We note that, if $\hat{O}$ is hermitian and $f$ a real function, the r.h.s. of the previous line is antisymmetric with respect to interchanging $\tau\leftrightarrow\tau'$ and hence becomes trivial.
 For nonhermitian operators, that is not necessarily the case.
 Equation \eqref{Eq:FDI} is a slightly generalized form of the fluctuation--dissipation inequality put forward in ref. \cite{fleming13}.
 If we now set $\hat{O}=\hat{\xi}$, Equation \eqref{Eq:FDI} simply states that
 the hermicity of $\hat{\xi}$ translates into a correlation function $\nu$ that is a positive semidefinite function in the sense of Equation \eqref{Eq:DefPosDefinite}.
 Alternatively, setting $f(\tau)=1$, substituting $x=\tau-\tau'$ and using
 \begin{align}
 \int_0^t\mathrm{d}\tau\int_{\tau-t}^{\tau}
 	&=
 	\int_{-t}^0\mathrm{d}x
 	\int_0^{t+x}
 	\mathrm{d}\tau
 	+
 	\int_0^t\mathrm{d}x
 	\int_x^t\mathrm{d}\tau
 ,
 \end{align}
 We find that the fluctuations integrated over the full span of time always are equal to or exceed the corresponding first moment of the time delay, i.e.,
 \begin{align}\label{Eq:TimeAvg}
 \int_0^{t}\mathrm{d}\tau~\nu(\tau)
 	&\geq
 	\frac{\int_0^{t}\mathrm{d}\tau~\tau\nu(\tau)}{t}
 .
 \end{align}
 Additionally, we can make a statement on the relation between $\nu$ and $\gamma$ in frequency domain.
 To this end, without loss of generality, we extend the integral boundaries in our definition of the scalar product in Equation \eqref{Eq:DefPosDefinite} to $\pm\infty$.
 It is then possible to show that for stationary noise and any instance of time, arbitrarily far from equilibrium, the noise spectrum of the environment is always equal to or exceeds the spectrum of the dissipation kernel weighted by the factor $\hbar\omega$ (see ref. \cite{fleming13} for details)
 which is stated in Equation \eqref{Eq:FDIFrequ}.
 We note that the generalization to multivariate noise is straight-forward \cite{fleming13,reiche20a}.
 Equation~\eqref{Eq:TimeAvg} explicitly shows that detailed balance is not necessarily fulfilled in nonequilibrium.


 \subsection{Robertson--Schrödinger Inequality}\label{Sec:RSI}

 In terms of the covariance matrix $\langle\Delta\mathbf{Q}^2\rangle_{\mathrm{s}}$, the Robertson--Schr\"odinger inequality can be stated as
 \begin{align}
 \label{Eq:RSUncertainty}
 \text{det}\langle\Delta\mathbf{Q}^2\rangle_{\mathrm{s}}
 	&=\det\left[\underline{\sigma}_0(t)+\underline{\sigma}(t)\right]
 	\geq
   \frac{1}{4}
   \left|\langle[\hat{q},\dot{\hat{q}}]\rangle\right|^2
 	.
 \end{align}
 We can specify the previous line to our specific situation.
 First, we note that the cross correlations can be written as
 \begin{subequations}
 \label{Eq:CrossCorr}
 \begin{align}
 \label{Eq:CrossCorr1}
 \langle(\hat{q}\dot{\hat{q}}-\langle\hat{q}\rangle\langle\dot{\hat{q}}\rangle)\rangle_{\mathrm{s}}
   &=
   [\underline{\sigma}_0]_{q\dot{q}}
   \\\nonumber
   &
   +
   \int_0^t\mathrm{d}\tau\int_0^t\mathrm{d}\bar{\tau}~
   \chi(\tau)\dot{\chi}(\bar{\tau})
   \nu(\tau-\bar{\tau}),
   \\
   \label{Eq:CrossCorr2}
 \langle[\hat{q},\dot{\hat{q}}]\rangle
   &
   =
   i\hbar \omega_0^2~
   \text{det}
   \left[
   \underline{X}(t)
   \right]
   \\\nonumber
   &
   +
   2i\hbar
   \int_0^t\mathrm{d}\tau
   \int_0^t\mathrm{d}\bar{\tau}~
   \chi(\tau)
   \dot{\chi}(\bar{\tau})
   \Gamma(\tau-\bar{\tau}),
 \end{align}
 \end{subequations}
 where the subscript denote the corresponding components of the matrix and we used that, at initial time $t=0$, the system and environment fulfill the usual canonical commutation relations $[\hat{q}(0),\dot{\hat{q}}(0)/\omega_0^2]=i\hbar$.
 Here, the extra $\omega_0^{-2}$ is due to the proper definition of the canonical momentum.
 Combining Equations \eqref{Eq:RSUncertainty} and \eqref{Eq:CrossCorr}, we obtain the time-dependent Robertson--Schrödinger function for strongly coupled system--bath dynamics 
 \begin{align}
 \label{Eq:RSUncertainty2}
 \langle\Delta &\hat{q}^2(t)\rangle_{\mathrm{s}}
 \langle\Delta \dot{\hat{q}}^2(t)\rangle_{\mathrm{s}}
   \geq
   \left|
   \langle(\hat{q}\dot{\hat{q}}-\langle\hat{q}\rangle\langle\dot{\hat{q}}\rangle)\rangle_{\mathrm{s}}
   \right|^2
   +
   \frac{1}{4}
   \left|\langle[\hat{q},\dot{\hat{q}}]\rangle\right|^2
   \\\nonumber
   &\geq
   \left(
   [\underline{\sigma}_0]_{q\dot{q}}
   +
   \int_0^t\mathrm{d}\tau\int_0^t\mathrm{d}\bar{\tau}~
   \chi(\tau)\dot{\chi}(\bar{\tau})\nu(\tau-\bar{\tau})
   \right)^2
   \\\nonumber
   &
   +\frac{\hbar^2\omega_0^4}{4}\text{det}[\underline{X}(t)]^2
   \\\nonumber
   &
   +
   \frac{\hbar^2\omega_0^2}{2}\text{det}[\underline{X}(t)]
   \int_0^t\mathrm{d}\tau
   \int_0^t\mathrm{d}\bar{\tau}~
   \chi(\tau)
   \dot{\chi}(\bar{\tau})
   \Gamma(\tau-\bar{\tau})
   \\\nonumber
   &+
   \hbar^2
   \left[
   \int_0^t\mathrm{d}\tau
   \int_0^t\mathrm{d}\bar{\tau}~
   \chi(\tau)
   \dot{\chi}(\bar{\tau})
   \Gamma(\tau-\bar{\tau})
   \right]^2.
 \end{align}
 It is quite interesting to consider the late-time limit of the previous inequality.
 To this end, we note that almost all components of the first three terms of Equation \eqref{Eq:RSUncertainty2} are given by elements of the linear response tensor $\underline{X}(t)$. Since the latter obeys the damped Equation~\eqref{Eq:GreenHomII}, it can generally be expected to vanish in the limit $t\to\infty$.
 The remaining term
 $\lim_{t\to\infty}
 \int_0^t\mathrm{d}\tau\int_0^t\mathrm{d}\bar{\tau}~
 \chi(\tau)\dot{\chi}(\bar{\tau})\nu(\tau-\bar{\tau})
 =0
 $
 also vanishes since we integrate an odd function over a symmetric integral.
 Only the last term in Equation \eqref{Eq:RSUncertainty2} prevails at late times, such that the Robertson--Schrödinger equation evolves according to
 \begin{align}
 \label{Eq:RSUncertainty3}
 \lim_{t\to\infty}\langle&\Delta \hat{q}^2(t)\rangle_{\mathrm{s}}
 \langle\Delta \dot{\hat{q}}^2(t)\rangle_{\mathrm{s}}
\\\nonumber
   &\geq
   \hbar^2
   \left[
   \int\mathrm{d}\tau
   \int\mathrm{d}\bar{\tau}~
   \chi(\tau)
   \dot{\chi}(\bar{\tau})
   \Gamma(\tau-\bar{\tau})
   \right]^2
   \\\nonumber
     &=
     \hbar^2
     \left[
     \int\frac{\mathrm{d}\omega}{2\pi}~
     \omega^2
     |\alpha(\omega)|^2
     \gamma(\omega)
     \right]^2.
 \end{align}
 For comparison, at late times, the fluctuation--dissipation inequality applied to the variance of position and momentum operators directly yields
 \begin{align}
 \label{Eq:RSUncertaintyFDI}
 &\lim_{t\to\infty}\langle\Delta \hat{q}^2(t)\rangle_{\mathrm{s}}
 \langle\Delta \dot{\hat{q}}^2(t)\rangle_{\mathrm{s}}
 \\\nonumber
   &=
   \left[
   \int\frac{\mathrm{d}\omega}{2\pi}~
   |\alpha(\omega)|^2
   \nu(\omega)
   \right]
   \left[
   \int\frac{\mathrm{d}\omega}{2\pi}~
   \omega^2
   |\alpha(\omega)|^2
   \nu(\omega)
   \right]
   \\\nonumber
     &\geq
     \hbar^2
     \left[
     \int\frac{\mathrm{d}\omega}{2\pi}~
     |\alpha(\omega)|^2
     |\omega\gamma(\omega)|
     \right]
     \left[
     \int\frac{\mathrm{d}\omega}{2\pi}~
     \omega^2
     |\alpha(\omega)|^2
     |\omega\gamma(\omega)|
     \right].
 \end{align}
 The connection between the fluctuation--dissipation inequality (FDI) and the Robertson--Schrödinger (RS) uncertainty relation is provided by the Cauchy--Schwarz (CS) inequality for integrals \cite{dlmf}, which further bounds Equation \eqref{Eq:RSUncertaintyFDI} as
 \begin{align}
 \label{Eq:FDIsConnected}
 &\lim_{t\to\infty}\langle\Delta \hat{q}^2(t)\rangle_{\mathrm{s}}
 \langle\Delta \dot{\hat{q}}^2(t)\rangle_{\mathrm{s}}
 \\\nonumber
 &\overset{\mathrm{FDI}}{\geq}
 \hbar^2\left[
 \int\frac{\mathrm{d}\omega}{2\pi}~
 |\alpha(\omega)|^2
 |\omega\gamma(\omega)|
 \right]
 \left[
 \int\frac{\mathrm{d}\omega}{2\pi}~
 \omega^2
 |\alpha(\omega)|^2
 |\omega\gamma(\omega)|
 \right]
   \\\nonumber
   &\overset{\mathrm{CS}}{\geq}
   \hbar^2\left[
   \int\frac{\mathrm{d}\omega}{2\pi}~
   \omega^2
   |\alpha(\omega)|^2
   |\gamma(\omega)|
   \right]^2
   \quad (\text{RS}).
 \end{align}
 In fact, it can be shown that the last line of the previous equation, i.e., the lower bound of the Robertson--Schr\"odinger equation, reproduces the conventional Heisenberg bound, since
 $
 \int\frac{\mathrm{d}\omega}{2\pi}~
 \omega^2
 |\alpha(\omega)|^2
 |\gamma(\omega)|
 =1/2
 $ \cite{dong22}.
 We kept the frequency representation as it makes the connection to the fluctuation--dissipation inequality more transparent.
 In this way, at late times, the fluctuation--dissipation inequality not only reproduces the Robertson--Schrödinger equality as a special case, but actually provides a stronger bound.

 From the previous uncertainty relations, especially the FDI in the second line of Equation \eqref{Eq:FDIsConnected}, it becomes clear that the strong coupling between the system and its environment leads to fluctuations that can exceed the usual Heisenberg bound.
 Indeed, the Heisenberg uncertainty relation should be restored in the limit of vanishingly small system+bath coupling which is implicitly encoded in the magnitude of $\gamma$.
 To show this explicitly, we focus on the late-time limit for simplicity and note that $\gamma(\tau)$ is a real function that is even in its argument such that $\gamma(\omega)=2\text{Re}[\gamma_{\theta}(\omega)]$.
 In the limit of small coupling, we then have that
 \begin{align}
 \omega
 |\alpha(\omega)|^2&
 |\gamma(\omega)|
 \sim
   \pi
   \delta
   \left(
   \omega_0^2-\omega^2+2\omega\text{Im}[\gamma_{\theta}(\omega)]
   \right)^2.
 \end{align}
 Here, $2\omega\text{Im}[\gamma_{\theta}(\omega)]$ can be seen as system's {resonance frequency} due to the coupling with the environment.
 In leading order coupling, this is subleading with respect to $\omega_0$, so that we obtain
 \begin{align}
 \lim_{\substack{t\to\infty\\ \gamma\to 0}}\langle\Delta \hat{q}^2(t)\rangle_{\mathrm{s}}
 \langle\Delta \dot{\hat{q}}^2(t)\rangle_{\mathrm{s}}
   &\geq \frac{\hbar^2}{4}
   \left[
   1+
   \mathcal{O}(\gamma(\omega))
   \right].
 \end{align}
 This is nothing but the Heisenberg uncertainty relation, and we refer to, e.g., refs. \cite{hu93a,hu95} for further details as well as ref. \cite{dong22} for a finite-time version of Equation \eqref{Eq:FDIsConnected}.


 \section{Wick's Theorem for Thermal States}\label{App:WicksTheorem}

 In the perturbative treatment of in--out quantum field theory, the series is often expanded in terms of the vacuum $n$-point correlation functions of free fields, that is, the vacuum expectation values of the time-ordered product of $n$ free-field operators.
 The Wick's theorem states~\cite{wick50,peskin95} that the time-ordered product of the free-field operators can be expressed as the sum of the normal-ordered product and the permutation sums of the products of pair-wise contractions. When the state is the vacuum state of the free field, the normal-ordered term vanishes and each contraction gives a Feynman propagator. Then, the Wick's theorem implies that the vacuum $n$-point correlation functions of the free field can be expanded by the permutation sums of the products of the corresponding two-point Feynman propagators. Thus, the perturbative quantum field theory reduces to the knowledge of the two-point Feynman propagator.

 However, the thermal expectation value of the normal-ordered term in general does not vanish. So, we may wonder whether the thermal expectation values of the time-ordered product of $n$ free-field operators can be likewise expanded by the finite-temperature Feynman propagator. Various generalizations of the original Wick' theorem are formulated for finite-temperature field theory~\cite{evans96}, in--in nonequilibrium field theory~\cite{plimak11}, and  arbitrary initial state of the field operators~\cite{leeuwen12}.

 Here, we summarize the Wick's theorem for the thermal scalar field, proved in~\cite{evans96} for readers' convenience. First, consider the thermal expectation value,
 \begin{align}
 	&\operatorname{Tr}\Bigl\{\hat{\rho}_{\beta}\hat{\alpha}_{a_{1}}\cdots\hat{\alpha}_{a_{n}}\Bigr\}\,,&&\text{with} &\hat{\alpha}_{a_{i}}=\begin{cases}
 		\hat{a}^{\vphantom{\dagger}}_{\bm{k}_{i}}\,,&a_{i}=-\,,\\
 		\hat{a}^{\dagger}_{\bm{k}_{i}}\,,&a_{i}=+\,.
  \end{cases}
 \end{align}
 where $\hat{a}^{\vphantom{\dagger}}_{\bm{k}_{i}}$, $\hat{a}^{\dagger}_{\bm{k}_{i}}$ are standard annihilation and creation operators of the free quantum scalar field, and $\hat{\rho}_{\beta}$ is the thermal state of the field. The trace can be written as
 \begin{align}
 	\operatorname{Tr}\Bigl\{\hat{\rho}_{\beta}\hat{\alpha}_{a_{1}}\cdots\hat{\alpha}_{a_{n}}\Bigr\}
   &=
   \operatorname{Tr}\Bigl\{\hat{\rho}_{\beta}\bigl[\hat{\alpha}_{a_{1}},\hat{\alpha}_{a_{2}}\bigr]\hat{\alpha}_{a_{3}}\cdots\hat{\alpha}_{a_{n}}\Bigr\}
   \\\nonumber
   &\quad
   +
   \operatorname{Tr}\Bigl\{\hat{\rho}_{\beta}\hat{\alpha}_{a_{2}}\hat{\alpha}_{a_{1}}\hat{\alpha}_{a_{3}}\cdots\hat{\alpha}_{a_{n}}\Bigr\}
   \\\nonumber
 	&=
   \operatorname{Tr}\Bigl\{\hat{\rho}_{\beta}\bigl[\hat{\alpha}_{a_{1}},\hat{\alpha}_{a_{2}}\bigr]\hat{\alpha}_{a_{3}}\cdots\hat{\alpha}_{a_{n}}\Bigr\}
   \\\nonumber
   &\quad
   +
   \operatorname{Tr}\Bigl\{\hat{\rho}_{\beta}\hat{\alpha}_{a_{2}}\bigl[\hat{\alpha}_{a_{1}},\hat{\alpha}_{a_{3}}\bigr]\hat{\alpha}_{a_{4}}\cdots\hat{\alpha}_{a_{n}}\Bigr\}
   \\\nonumber
   &\quad
   +
   \cdots
   \\\nonumber
 	&\quad+\operatorname{Tr}\Bigl\{\hat{\rho}_{\beta}\hat{\alpha}_{a_{2}}\hat{\alpha}_{a_{3}}\cdots\hat{\alpha}_{a_{n-1}}\bigl[\hat{\alpha}_{a_{1}},\hat{\alpha}_{a_{n}}\bigr]\Bigr\}
   \\\nonumber
   &\quad
   +
   \operatorname{Tr}\Bigl\{\hat{\rho}_{\beta}\hat{\alpha}_{a_{2}}\hat{\alpha}_{a_{3}}\cdots\hat{\alpha}_{a_{n}}\hat{\alpha}_{a_{1}}\Bigr\}\,.\label{E:fkjdgf}
 \end{align}
 By the Baker--Campbell--Hausdorff formula, we have
 \begin{align}
 	e^{\beta\hat{H}}\hat{a}\,e^{-\beta\hat{H}}&=e^{-\beta\omega}\,\hat{a}\,,&e^{\beta\hat{H}}\hat{a}^{\dagger}e^{-\beta\hat{H}}&=e^{+\beta\omega}\,\hat{a}^{\dagger}\,,
 \end{align}
 so, we can write the last term in \eqref{E:fkjdgf} as
 \begin{align}
 	\operatorname{Tr}&\Bigl\{\hat{\rho}_{\beta}\hat{\alpha}_{a_{2}}\hat{\alpha}_{a_{3}}\cdots\hat{\alpha}_{a_{n}}\hat{\alpha}_{a_{1}}\Bigr\}
   \\\nonumber
   &
   =
   e^{-\lambda_{a_{1}}\beta\omega_{\bm{k}_{1}}}\operatorname{Tr}\Bigl\{\hat{\rho}_{\beta}\hat{\alpha}_{a_{1}}\hat{\alpha}_{a_{2}}\hat{\alpha}_{a_{3}}\cdots\hat{\alpha}_{a_{n}}\Bigr\}\,,
 \end{align}
 where $\lambda_{-}=+1$ and $\lambda_{+}=-1$.
 Thus, we find
 \begin{align}
 	\operatorname{Tr}\Bigl\{
   \hat{\rho}_{\beta}
   &
   \hat{\alpha}_{a_{1}}
   \cdots\hat{\alpha}_{a_{n}}\Bigr\}
   =
   \frac{\bigl[\hat{\alpha}_{a_{1}},\hat{\alpha}_{a_{2}}\bigr]}{1-e^{-\lambda_{a_{1}}\beta\omega_{\bm{k}_{1}}}}\,\operatorname{Tr}\Bigl\{\hat{\rho}_{\beta}\hat{\alpha}_{a_{3}}\cdots\hat{\alpha}_{a_{n}}\Bigr\}
   \\\nonumber
   &
   \\\nonumber
   &
   +\frac{\bigl[\hat{\alpha}_{a_{1}},\hat{\alpha}_{a_{3}}\bigr]}{1-e^{-\lambda_{a_{1}}\beta\omega_{\bm{k}_{1}}}}\,\operatorname{Tr}\Bigl\{\hat{\rho}_{\beta}\hat{\alpha}_{a_{2}}\hat{\rho}_{\beta}\hat{\alpha}_{a_{4}}\cdots\hat{\alpha}_{a_{n}}\Bigr\}
   +\cdots
   \\\nonumber
   &
   +
   \frac{\bigl[\hat{\alpha}_{a_{1}},\hat{\alpha}_{a_{n}}\bigr]}{1-e^{-\lambda_{a_{1}}\beta\omega_{\bm{k}_{1}}}}\,\operatorname{Tr}\Bigl\{\hat{\rho}_{\beta}\hat{\alpha}_{a_{2}}\hat{\rho}_{\beta}\hat{\alpha}_{a_{4}}\cdots\hat{\alpha}_{a_{n-1}}\Bigr\}\,.
 \end{align}
 We next rewrite the factor of the form
 $
 	\bigl[\hat{\alpha}_{a_{1}},\hat{\alpha}_{a_{2}}\bigr]
   (1-e^{-\lambda_{a_{1}}\beta\omega_{\bm{k}_{1}}})^{-1}\,.
 $
 If $a_{1}=-$, that is, $\hat{\alpha}_{a_{1}}=\hat{a}_{\bm{k}_{1}}$, then the factor is possibly nonzero only if $a_{2}=+$, i.e.,  $\hat{\alpha}_{a_{2}}=\hat{a}^{\dagger}_{\bm{k}_{2}}$, and we obtain ($a_1=-$, $a_2=+$)
 \begin{align}
   \frac{\bigl[\hat{\alpha}_{a_{1}},\hat{\alpha}_{a_{2}}\bigr]}{1-e^{-\lambda_{a_{1}}\beta\omega_{\bm{k}_{1}}}}
   &=\delta_{\bm{k}_{1}\bm{k}_{2}}\,\frac{1}{1-e^{-\beta\omega_{\bm{k}_{1}}}}
   \\\nonumber
   &
   =
   \delta_{\bm{k}_{1}\bm{k}_{2}}\,\bigl(N_{\bm{k}_{1}}+1\bigr)
   \\\nonumber
   &
   =
   \langle\hat{a}^{\vphantom{\dagger}}_{\bm{k}_{1}}\hat{a}^{\dagger}_{\bm{k}_{2}}\rangle_{\beta}\,
 \end{align}
with $N_{\bm{k}_1}$ the Bose occupation number at frequency $\omega_{\bm{k}_1}$ and $\langle\cdot\rangle_{\beta}$ the average over a thermal density matrix at inverse temperature $\beta$.
 On the other hand, if $a_{1}=+$, that is, $\hat{\alpha}_{a_{1}}=\hat{a}^{\dagger}_{\bm{k}_{1}}$, then the factor is possibly nonzero only if $a_{2}=-$, i.e.,  $\hat{\alpha}_{a_{2}}=\hat{a}^{\vphantom{\dagger}}_{\bm{k}_{2}}$, and we have
 ($a_1=+$, $a_2=-$)
 \begin{align}
 \frac{\bigl[\hat{\alpha}_{a_{1}},\hat{\alpha}_{a_{2}}\bigr]}{1-e^{-\lambda_{a_{1}}\beta\omega_{\bm{k}_{1}}}}
 &=
 -\delta_{\bm{k}_{1}\bm{k}_{2}}\,\frac{1}{1-e^{\beta\omega_{\bm{k}_{1}}}}
 \\\nonumber
 &
 =
 \delta_{\bm{k}_{1}\bm{k}_{2}}\,N_{\bm{k}_{1}}
 \\\nonumber
 &
 =\langle\hat{a}^{\dagger}_{\bm{k}_{1}}\hat{a}^{\vphantom{\dagger}}_{\bm{k}_{2}}\rangle_{\beta}\,.
 \end{align}
 Thus, altogether, we arrive at
 \begin{equation}
 	\frac{\bigl[\hat{\alpha}_{a_{1}},\hat{\alpha}_{a_{2}}\bigr]}{1-e^{-\lambda_{a_{1}}\beta\omega_{\bm{k}_{1}}}}
   =
   \delta_{a_{1}-}\delta_{a_{2}+}\,\langle\hat{a}^{\vphantom{\dagger}}_{\bm{k}_{1}}\hat{a}^{\dagger}_{\bm{k}_{2}}\rangle_{\beta}
   +\delta_{a_{1}+}\delta_{a_{2}-}\,\langle\hat{a}^{\dagger}_{\bm{k}_{1}}\hat{a}^{\vphantom{\dagger}}_{\bm{k}_{2}}\rangle_{\beta}\,.
 \end{equation}
 With the notations we introduce here, we can write a real scalar field operator as
 \begin{equation}
 	\hat{\phi}(x)=\sum_{\bm{k}}\sum_{a=\pm}\hat{\alpha}_{a}f_{a}(x)
 \end{equation}
 with
 \begin{align}
 	\hat{\alpha}_{a}&=\begin{cases}
 						\hat{a}_{\bm{k}}\,,&a=-\,,\\
 						\hat{a}^{\dagger}_{\bm{k}}\,,&a=+\,,
 	\end{cases}
 	&f_{a}(x)&=\begin{cases}
 						u_{\bm{k}}\,,&a=-\,,\\
 						u^{*}_{\bm{k}}\,,&a=+\,.
 	\end{cases}
 \end{align}
 where $u_{\bm{k}}$ is the mode function. Therefore, we can write
 \begin{equation}
 	\mathfrak{W}_{123\cdots n}=\langle\hat{\phi}(x_{1})\cdots\hat{\phi}(x_{n})\rangle_{\beta}
 \end{equation}
 as
 \begin{widetext}
 \begin{align}
 	\mathfrak{W}_{123\cdots n}&=\sum_{\bm{k}_{1}\cdots\bm{k}_{n}}\sum_{a_{1}\cdots a_{n}}\operatorname{Tr}\Bigl\{\hat{\rho}_{\beta}\hat{\alpha}_{a_{1}}\cdots\hat{\alpha}_{a_{n}}\Bigr\}\,f_{a_{1}}(x_{1})\cdots f_{a_{n}}(x_{n})\notag\\
 	&=\sum_{\bm{k}_{1}\bm{k}_{2}}\sum_{a_{1}a_{2}}\Bigl[\delta_{a_{1}-}\delta_{a_{2}+}\,\langle\hat{a}^{\vphantom{\dagger}}_{\bm{k}_{1}}\hat{a}^{\dagger}_{\bm{k}_{2}}\rangle_{\beta}\,u^{\vphantom{*}}_{\bm{k}_{1}}(x_{1})u^{*}_{\bm{k}_{2}}(x_{2})+\delta_{a_{1}+}\delta_{a_{2}-}\,\langle\hat{a}^{\dagger}_{\bm{k}_{1}}\hat{a}^{\vphantom{\dagger}}_{\bm{k}_{2}}\rangle_{\beta}\,u^{*}_{\bm{k}_{1}}(x_{1})u^{\vphantom{*}}_{\bm{k}_{2}}(x_{2})\Bigr]\notag\\
 	&\qquad\qquad\times\sum_{\bm{k}_{3}\cdots\bm{k}_{n}}\sum_{a_{3}\cdots a_{n}}\operatorname{Tr}\Bigl\{\hat{\rho}_{\beta}\hat{\alpha}_{a_{3}}\cdots\hat{\alpha}_{a_{n}}\Bigr\}\,f_{a_{3}}(x_{3})\cdots f_{a_{n}}(x_{n})+\cdots\notag\\
 	&=\langle\hat{\phi}(x_{1})\hat{\phi}(x_{2})\rangle_{\beta}\sum_{\bm{k}_{3}\cdots\bm{k}_{n}}\sum_{a_{3}\cdots a_{n}}\operatorname{Tr}\Bigl\{\hat{\rho}_{\beta}\hat{\alpha}_{a_{3}}\cdots\hat{\alpha}_{a_{n}}\Bigr\}\,f_{a_{3}}(x_{3})\cdots f_{a_{n}}(x_{n})\notag\\
 	&\qquad+\langle\hat{\phi}(x_{1})\hat{\phi}(x_{3})\rangle_{\beta}\sum_{\bm{k}_{2}\bm{k}_{4}\cdots\bm{k}_{n}}\sum_{a_{2}a_{4}\cdots a_{n}}\operatorname{Tr}\Bigl\{\hat{\rho}_{\beta}\hat{\alpha}_{a_{2}}\hat{\alpha}_{a_{4}}\cdots\hat{\alpha}_{a_{n}}\Bigr\}\,f_{a_{2}}(x_{2})f_{a_{4}}(x_{4})\cdots f_{a_{n}}(x_{n})+\cdots\,.
 \end{align}
 \end{widetext}
 for all permutations. Here $\langle\hat{\phi}(x_{1})\hat{\phi}(x_{2})\rangle_{\beta}$, for example, is the thermal Feynman propagator, constructed from the field operator evaluated at the spacetime points $x_{1}$ and $x_{2}$.
 In particular, the four-point function $\mathfrak{W}_{1234}=\langle\hat{\phi}(x_{1})\hat{\phi}(x_{2})\hat{\phi}(x_{3})\hat{\phi}(x_{4})\rangle_{\beta}$ can be expanded as
 \begin{align}\label{E:rtbgdgsdf}
 	\langle\hat{\phi}(x_{1})\hat{\phi}(x_{2})\hat{\phi}(x_{3})\hat{\phi}(x_{4})\rangle_{\beta}
   &=
   \langle\hat{\phi}(x_{1})\hat{\phi}(x_{2})\rangle_{\beta}\langle\hat{\phi}(x_{3})\hat{\phi}(x_{4})\rangle_{\beta}
   \nonumber\\
   &+
   \langle\hat{\phi}(x_{1})\hat{\phi}(x_{3})\rangle_{\beta}\langle\hat{\phi}(x_{2})\hat{\phi}(x_{4})\rangle_{\beta}
   \nonumber\\
 	&+
   \langle\hat{\phi}(x_{1})\hat{\phi}(x_{4})\rangle_{\beta}\langle\hat{\phi}(x_{2})\hat{\phi}(x_{3})\rangle_{\beta}\,.
 \end{align}
 Note that the ordering of the operators matters.
 Equation~\eqref{E:rtbgdgsdf} implies that we can write
 \begin{align}\label{E:gbsswr}
 &	\langle\bigl\{
   \hat{\phi}(x_{1})
   ,\hat{\phi}(x_{2})\bigr\}
   \bigl\{\hat{\phi}(x_{3}),\hat{\phi}(x_{4})\bigr\}\rangle_{\beta}
   \nonumber\\
   &=
   \mathfrak{W}_{1234}+\mathfrak{W}_{2134}+\mathfrak{W}_{1243}+\mathfrak{W}_{2143}
   \nonumber\\
 	&=\mathfrak{W}_{12}\mathfrak{W}_{34}+\mathfrak{W}_{13}\mathfrak{W}_{24}+\mathfrak{W}_{14}\mathfrak{W}_{23}+\mathfrak{W}_{21}\mathfrak{W}_{34}
   \nonumber\\
 	&\quad+\mathfrak{W}_{23}\mathfrak{W}_{14}+\mathfrak{W}_{24}\mathfrak{W}_{13}+\mathfrak{W}_{12}\mathfrak{W}_{43}+\mathfrak{W}_{14}\mathfrak{W}_{23}
   \nonumber\\
   &\quad+\mathfrak{W}_{13}\mathfrak{W}_{24}+\mathfrak{W}_{21}\mathfrak{W}_{43}+\mathfrak{W}_{24}\mathfrak{W}_{13}+\mathfrak{W}_{23}\mathfrak{W}_{14}
   \nonumber\\
 	&=\langle\bigl\{\hat{\phi}(x_{1}),\hat{\phi}(x_{2})\bigr\}\rangle_{\beta}\langle\bigl\{\hat{\phi}(x_{3}),\hat{\phi}(x_{4})\bigr\}\rangle_{\beta}
   \nonumber\\
   &\quad+
   4\mathfrak{W}_{13}\mathfrak{W}_{24}+4\mathfrak{W}_{14}\mathfrak{W}_{23}\,.
 \end{align}
 This allows us to compute the energy current fluctuations.

 \section{Fluctuating Energy Flow}
 \label{App:CalcFlucEnergyFlow}

 We provide the necessary steps to derive Equation \eqref{Eq:powerFluct}, i.e., connecting the derivative of the mechanical energy stores in the operator $\hat{Q}$ explicitly with the incoming and the outgoing power.
 The nontrivial part of the relation is to connect the expression for $P_{\rm out}^{\rm fluc}$ in Equation \eqref{Eq:PoutTimeDomain} to the form used in Equation \eqref{Eq:powerFluct}.
 We start from Equation \eqref{Eq:PoutTimeDomain} and note that the threefold integral can be partially decoupled using
 \begin{align}
 \label{Eq:AuxInt2}
 \int_0^t\mathrm{d}\tau\int_0^{\tau}\mathrm{d}x=\int_0^t\mathrm{d}x\int_x^t\mathrm{d}\tau
 \end{align}
 such that we find
 \begin{align}
 &\int_0^t\mathrm{d}\tau~
 	\gamma(t-\tau)
 	\int_0^{\tau}\mathrm{d}\tau'~
 	\dot{\chi}(\tau-\tau')
 	f(t,\tau')
 	\\\nonumber
 	&
 	=
 	\int_0^{t}\mathrm{d}\tau'~
 	\int_{\tau'}^t\mathrm{d}\tau~
 	\gamma(t-\tau)
 	\dot{\chi}(\tau-\tau')
 	f(t,\tau')
 	\\\nonumber
 	\overset{x=\tau-\tau'}{=}
   &
 	\int_0^{t}\mathrm{d}\tau'~
 	\left[
 	\int_{0}^{t-\tau'}\mathrm{d}x~
 	\gamma([t-\tau']-x)
 	\dot{\chi}(x)
 	\right]
 	f(t,\tau'),
 \end{align}
 as well as
 \begin{align}
 &\int_0^{t}\mathrm{d}\tau'~
 \int_0^t\mathrm{d}\tau''~
 	g(t-\tau')
 	\dot{\chi}(t-\tau'')
 	\nu(\tau'-\tau'')
 	\\\nonumber
 	&
   \quad
 	=
 	\int_0^t\mathrm{d}y~
 	\dot{\chi}(y)
 	\int_0^t\mathrm{d}z~
 	g(z)
 	\nu(y-z).
 \end{align}
 for arbitrary functions $f,g$.
 Combining the previous lines yields
 \begin{align}
 &\int_0^t\mathrm{d}\tau
 	\gamma(t-\tau)
 	\int_0^{\tau}\mathrm{d}\tau'~
 	\dot{\chi}(\tau-\tau')
 	\\\nonumber
 	&\quad\times
   \int_0^{t}\mathrm{d}\tau''~
 	\dot{\chi}(t-\tau'')
 	\nu(\tau'-\tau'')
 	\\\nonumber
 	&\quad
 	=
 	\int_0^t\mathrm{d}y~
 	\dot{\chi}(y)
 	\int_0^t\mathrm{d}z~
 	\left[
 	\int_{0}^{z}\mathrm{d}x~
 	\gamma(z-x)
 	\dot{\chi}(x)
 	\right]
 	\nu(y-z)
 	\\\nonumber
 	&\quad
 	=
 	-
 	\int_0^t\mathrm{d}y~
 	\dot{\chi}(y)
 	\int_0^t\mathrm{d}z~
 	\left[
 	\ddot{\chi}(z)+\omega_0^2\chi(z)
 	\right]
 	\nu(y-z).
 \end{align}
 The latter reproduces Equation \eqref{Eq:powerFluct}.

 \section{Master Equation, Density Matrix, and Von Neumann Entropy}
 \label{App:DensityMatrix}

 Following refs. \cite{calzetta03,fleming11} (see also Refs. \cite{halliwell96,calzetta01,ford01,intravaia03,fleming11,pagel13} for details), it is possible to connect the solution to the quantum Langevin Equation \eqref{Eq:qSol} to a master equation for the corresponding Wigner function $\mathcal{W}_r(q,\dot{q},t)$ ($q,\dot{q}$ are now to be understood as coordinates and not operators).
 To this end, we start from the statistical relation
 \begin{align}
 \label{Eq:FeynmanDelta}
 \mathcal{W}_r(q,\dot{q},t)
 	&=
 	\langle\langle
 	\delta[\mathbf{Q}(t)-\mathbf{Q}]\rangle_{\xi}\rangle_{\mathbf{Q}_0},
 \end{align}
 where we average over the initial conditions $\mathbf{Q}_0=\mathbf{Q}(0)$ (see Equation \eqref{Eq:qSol}) as well as the impact of the environment by means of the Feynman--Vernon functional integral (see refs. \cite{feynman63,calzetta03,fleming11} for details).
 Equation \eqref{Eq:FeynmanDelta} yields for the characteristic function (Wigner function in Fourier domain $\mathbf{Q}\leftrightarrow\mathbf{k}$)
 \begin{subequations}
 \begin{align}
 \mathcal{W}_r(\mathbf{k},t)
 	&=
 	\mathcal{W}_r\left(\underline{X}^T\mathbf{k},0\right)
 	e^{-\frac{1}{2}\mathbf{k}^T\underline{\sigma}(t)\mathbf{k}},
 \end{align}
 \end{subequations}
 where we use the matrix {$\underline{X}$}
  defined in Equation \eqref{Eq:DefX}.
 Upon differentiation with respect to time and performing a Fourier transform, we obtain the master equation
 \begin{align}
 \label{Eq:HuPazZhang}
 \dot{\mathcal{W}_r}&(q,\dot{q},t)
 	=
 	\\\nonumber
 	&
   \left\{
 	\nabla^T_Q\underline{\Phi}\mathbf{Q}
 	+
 	\frac{1}{2}
 	\nabla^T_Q
 	\left[
 	\underline{\Phi}\underline{\sigma}
 	+
 	\underline{\sigma}
 	\underline{\Phi}^T
 	+
 	\dot{\underline{\sigma}}
 	\right]
 	\nabla_Q
 	\right\}
 	\mathcal{W}_r(q,\dot{q},t),
 \end{align}
 where we introduced the matrix
 $
 \underline{\Phi}
   =-\dot{\underline{X}}\underline{X}
 $
 for convenience of notation and the gradient with respect to the coordinates, $\mathbf{Q}$ is to be understood as
 $
 \nabla_{Q}
   \equiv (\partial_q,\partial_{\dot{q}})^T
 $.
 Equation \eqref{Eq:HuPazZhang} is the Hu--Paz--Zhang master equation with time-dependent coefficients \cite{hu92}.
 We note that a direct connection to our initial quantum Langevin equation  [Equation \eqref{Eq:QLE}] can be seen from recasting the latter in its time-local form \cite{karrlein97,ford01}, i.e.,
 \begin{align}
 \ddot{\hat{q}}+\text{Tr}[\underline{\Phi}]\dot{\hat{q}}+\det[\underline{\Phi}]\hat{q}=\hat{\xi}(t).
 \end{align}

 For given $\chi(t)$ and $\nu(t)$ (specifics of the system), the dynamics of the system in quasi-phase space is determined by the initial conditions as well as the environment-induced {\textit{Gaussian}}
  covariance matrix.
 Choosing a simple Gaussian form for the initial Wigner~function
 \begin{align}
 \mathcal{W}_r(\underline{X}^T\mathbf{k},0)
 	&=
 	e^{-\frac{1}{2}\mathbf{k}^T\underline{\sigma}_0(t)\mathbf{k}-i\mathbf{k}\langle\mathbf{Q}(t)\rangle},
 \end{align}
 We can perform the Fourier transform exactly and obtain
 \begin{align}
 \label{Eq:WignerFunction}
 \mathcal{W}_r(\mathbf{Q},t)
 	&=
 	\int\frac{\mathrm{d}^2\mathbf{k}}{(2\pi)^2}~
 	e^{-\frac{1}{2}\mathbf{k}\left[\underline{\sigma}_0(t)+\underline{\sigma}(t)\right]\mathbf{k}}
 	e^{i\mathbf{k}[\mathbf{Q}-\langle\mathbf{Q}(t)\rangle]}
 	\\\nonumber
 	&=
 	\frac{1}{2\pi}
 	\frac{e^{-\frac{1}{2}
 		(\mathbf{Q}-\langle \mathbf{Q}(t)\rangle)^T
 		\left[\underline{\sigma}_0(t)+\underline{\sigma}(t)\right]^{-1}
 		(\mathbf{Q}-\langle \mathbf{Q}(t)\rangle)
 		}}{\sqrt{|\det\left[\underline{\sigma}_0(t)+\underline{\sigma}(t)\right]|}}.
 \end{align}
 Note that $\int\mathrm{d}^2\mathbf{Q}~\mathcal{W}_r(\mathbf{Q},t)=1$.
 Finally, for the density matrix in position representation, we need to perform yet another series of Gaussian integrals and obtain
 \begin{align}
 &\rho(q-q'/2,q+q'/2)
 	=
 	\int\mathrm{d}\dot{q}~
 	e^{-i \dot{q}q'}\mathcal{W}_r(q,\dot{q},t)
 	\\\nonumber
 	&=
 	\frac{1}{\sqrt{2\pi}}
 	\frac{1}{\sqrt{|[\tilde{\underline{\sigma}}(t)]_{qq}|}}
 	\\\nonumber
 	&\times
 	e^{-\frac{1}{2}
 		\frac{(q-\langle \hat{q}(t)\rangle)^2+\det[\tilde{\underline{\sigma}}(t)]q'^2+2i \left([\tilde{\underline{\sigma}}(t)]_{qq}\langle \dot{q}(t)\rangle+[\tilde{\underline{\sigma}}(t)]_{q\dot{q}}\{q-\langle \hat{q}(t)\rangle\}q'\right)}
 		{[\tilde{\underline{\sigma}}(t)]_{qq}}},
 \end{align}
 where we have defined
 $
 \underline{\tilde{\sigma}}(t)
 \equiv
   \underline{\sigma}_0(t)+\underline{\sigma}(t)
 $
 to shorten notation, and the subscript ``qq'' denotes the corresponding entry of the matrix.
 For vanishing mean values, i.e. $\langle \hat{q}(t)\rangle=\langle \dot{\hat{q}}(t)\rangle=0$, the previous line coincides with Equation (3.23) of ref. \cite{hsiang21}.
 Such a result is quite remarkable and was extensively discussed in refs. \cite{pagel13,hsiang21} because the density matrix $\hat{\rho}$ can now be rewritten in a Gibbs form
 \begin{align}
 \hat{\rho}
 	&=
 	\frac{1}{\mathcal{Z}}
 	e^{-\beta(t)\hat{H}(t)}
 \end{align}
 with effective temperature
 \begin{align}
 \label{Eq:BetaEff}
 \beta(t)
 	&=
 	\frac{2}{\hbar\omega_0}\sinh^{-1}
   \left[
   \frac{
   \left(
   \det\tilde{\underline{\sigma}}(t)-\frac{1}{4}
   \right)^{-\frac{1}{2}}}{2}
   \right]
 \end{align}
Quantifying the coupling between system and environment and the effective time-dependent Hamiltonian
 \begin{align}
 \hat{H}(t)
 	&=
 	\beta(t)^{-1}
 	\frac{\ln\frac{\det\tilde{\underline{\sigma}}(t)+\frac{1}{2}}{\det\tilde{\underline{\sigma}}(t)-\frac{1}{2}}}{2\sqrt{|\det\tilde{\underline{\sigma}}(t)|}}
 	\\\nonumber
 	&
   \times
 	\left[
 	\langle\hat{q}^2(t)\rangle \hat{q}^2
 	+
 	\langle\dot{\hat{q}}^2(t)\rangle\dot{\hat{q}}^2
 	-
 	\langle\hat{q}(t)\dot{\hat{q}}(t)\rangle_{\mathrm{s}}
 	\{\hat{q},\dot{\hat{q}}\}
 	\right].
 \end{align}
 Note that the Robertson--Schr\"odinger inequality is directly encoded in the temperature as well as in the corresponding partition function
 \begin{align}
 \mathcal{Z}=\sqrt{\det\tilde{\underline{\sigma}}(t)-\frac{1}{4}}.
 \end{align}
 Combining the previous results leads to Equation \eqref{Eq:Entropy} used in the main text \cite{hsiang21,colla21}.

 \section{Bath Spectral Density and Dissipation Kernel}
 \label{App:BathSpectrDens}

We review the common bath spectral densities and their analytic properties (see Equation~\eqref{Eq:SpectralDen} and discussion below).
 First, we note that we can write ($\omega_n=-\omega_{-n}^*$, $\omega_n=\omega_n^R-i\omega_n^I$)~\cite{Note2}.
 \begin{subequations}
 \begin{align}
 \Gamma_{\theta}(\omega)
   &=
   \frac{\omega}{2}\sum_n\frac{c_n}{\omega_n-\omega},
   \\
 I(\omega)
   &=
   \frac{\omega}{\pi}
   \sum_n
   \frac{
   c_n\omega_n^I
   }
   {(\omega-\omega_n^R)^2+(\omega_n^I)^2}
 .
 \end{align}
 \end{subequations}
 with complex constants $c_{-n}=c_n^*$.
 We can then evaluate the dissipation kernel by means of the residue theorem
 \begin{align}
 \gamma(\tau)
 	&
   =
   \sum_{n}
   \int\frac{\mathrm{d}\omega}{2\pi}~
   \frac{
   c_n\omega_n^I
   }
   {(\omega+ \omega_n^R)^2+(\omega_n^I)^2}
   e^{-i\omega\tau}
   \\\nonumber
   &
   =
   2
   \sum_{n>0}
   e^{-\omega_n^I|\tau|}
   \text{Re}
   \left[
   c_n
   e^{-i\omega_n^R\tau}
   \right]
   .
 \end{align}
 We note that, for simplicity, we have implicitly assumed that there are no connected branch cuts in the bath spectral density.

 Let us introduce two relevant examples for the bath spectral density.
 One convenient example that is often used is to have $I(\omega)$ either featuring exponential~\cite{hu95} or {Lorentzian damping} \cite{fleming11,einsiedler20} at the order of a cut-off frequency $\Lambda$. In the latter case, choosing so-called Ohmic damping (linear in frequency), we can write
 \begin{align}
 \label{Eq:ILambda}
 I_{\Lambda}(\omega)
   &=
   \frac{2\gamma_0}{\pi}
   \frac{\omega}{1+\frac{\omega^2}{\Lambda^2}}
 \end{align}
 with $\gamma_0>0$ a phenomenological damping rate.
 We can identify $\omega_n^R=0$, $\omega_n^I=\Lambda$, and $c_n=2\gamma_0\Lambda$, and hence find
 \begin{align}
 \label{Eq:LambdaGamma}
 \gamma_{\Lambda}(\tau)
   &=\gamma_0\Lambda e^{-\Lambda|\tau|}.
 \end{align}
 Another, more physical example, occurs when the system is coupled to a bath that features collective oscillations at frequency $\Omega$ and width $\kappa$, where
 \begin{align}
 \label{Eq:SpecDenOscillator}
 I_{\Omega}(\omega)
   &=
   \frac{\omega}{\pi}
   \frac{2\rho(\mathbf{r}_a)\kappa\Omega^2}{(\Omega^2-\omega^2)^2+\kappa^2\omega^2}.
 \end{align}
 Here, $\rho(\mathbf{r}_a)$ is a geometric factor that gives the strength of the interaction and could, in principle, depend on the position of the particle.
 For instance, if the system is an atom in the vicinity of a macroscopic object, $\rho(\mathbf{r}_a)$ would be connected to the small frequency value of the local density of states of the material-modified electromagnetic vacuum \cite{joulain03}. For positions $\mathbf{r}_a$ close to an interface, $\Omega$ would then be related to the surface plasmon- or  phonon-polariton frequency.
 See, e.g., refs. \cite{strasberg18,intravaia19a,klatt21,oelschlaeger22} for some applications of such a description.
 {Furthermore, we note that Equation \eqref{Eq:SpecDenOscillator} is substantially different from Equation~\eqref{Eq:ILambda}, as it features its very own damping mechanism described by the phenomenological dissipation rate $\kappa$ that is connected to the width of $\Omega$}.
 Although it is possible to solve the frequency integral for $\gamma(\tau)$ with the spectral bath density of Equation \eqref{Eq:SpecDenOscillator} by means of an exact partial fraction decomposition, it is simpler to consider the limit $\kappa\ll\Omega$, which holds for many realistic cases, i.e.,
 \begin{align}
 \label{Eq:SpecDensApprox}
 I_{\Omega}(\omega)
   \sim
   \frac{\omega\rho(\mathbf{r}_a)}{\pi}
   \left(
   \frac{\kappa/2}{(\frac{\kappa}{2})^2+(\omega-\Omega)^2}
   +
   \frac{\kappa/2}{(\frac{\kappa}{2})^2+(\omega+\Omega)^2}
   \right).
 \end{align}
 This shows four poles in the complex frequency plane at
 $
 \omega=\pm\Omega\pm i\kappa
 $,
 and we can identify $\omega_n^R=\pm\Omega$, $\omega_n^I=\kappa/2$ and $c_n=\rho_a(\mathbf{r}_a)$.
 Consequently, we obtain for the dissipation kernel
 \begin{align}
 \gamma_{\Omega}(\tau)
   &\sim
   \rho(\mathbf{r}_a)
   e^{-\frac{\kappa}{2}|\tau|}
   \cos[\Omega\tau].
 \end{align}
 The spectral features of the environment asymptotically show as an overall oscillatory behavior in the dissipation kernel.

 Lastly, it is relevant to mention that the bath spectral densities lead to an intrinsically non-Markovian dissipation kernel $\gamma$.
 Indeed, only in the limit of an infinitely large cut-off frequency $\Lambda$, the Ohmic damping turns to its (classical) Markovian value
 $\gamma_{\Lambda}(\tau)\to 2\gamma_0\delta(\tau)$.
 When the bath features at least one finite collective resonance, the situation is even more intricate, as the non-Markovianity remains even at zero damping, i.e., $\gamma_{\Omega}(\tau)\to\rho(\mathbf{r}_a)\cos[\Omega\tau]$ if $\kappa\to 0$.
 This non-Markovianity can only be removed by fully decoupling the oscillating behavior of system and environment, i.e. for all frequencies $\omega$ relevant for the interaction we have that
 $\lim_{\Omega\gg\omega}I_{\Omega}(\omega)
 $
 reproduces the Ohmic spectral density in Equation \eqref{Eq:ILambda} with $\gamma_0=\rho(\mathbf{r}_a)\kappa$ and $\Lambda=\Omega/\kappa$.
 For further details as well as methods for quantifying the degree of non-markovianity, we refer to, e.g., refs. \cite{strasberg18,das20,einsiedler20} (see also the reviews \cite{breuer16,vega17}).

 \section{Analytical and Numerical Treatment of the Correlation Functions}\label{App:IntNoiseKernel}

 Given the bath spectral densities in Appendix \ref{App:BathSpectrDens}, we compute analytical expressions for the noise kernel. We further give details on the numerical integration used for calculating correlation functions for the system degrees of freedom.

 From the expressions in Equation \eqref{Eq:NuGamma}, we can derive corrections to the inequality in Equation \eqref{Eq:ThermalInequality} up to arbitrary order.
 To this end, we employ the series expansion of the hyperbolic cotangent \cite{ford96c,kim07,estrada02}.
 \begin{align}
 \coth\left[\frac{\hbar\beta\omega}{2}\right]
 &=
 \frac{2}{\hbar\beta}
 \sum_{n=0}^{\infty}(2-\delta_{n0})
 \frac{\omega}{\omega^2+\left(\frac{2\pi n}{\hbar\beta}\right)^2}
 \end{align}
 and integrate the frequency integral for the noise correlation in every order $n$, i.e.,
 \begin{align}
 \label{Eq:NuIntegral}
 \nu(\tau)
   &=
   \frac{2}{\beta}
   \left(
   \gamma(\tau)
   +
   \sum_{n=1}^{\infty}
   \int\mathrm{d}\omega~
 	I(\omega)
   \frac{\omega e^{-i\omega\tau}}{\omega^2+(n\eta)^2}
   \right)
   ,
 \end{align}
 where $\eta=2\pi/\hbar\beta$ is the first Matsubara frequency \cite{matsubara55,reiche20}.
 We can solve the frequency integral quite generally by means of the residue theorem.
 For simplicity, we focus on isolated, simple  poles of the bath spectral density in the complex frequency plane such that the spectral density takes the form
 \begin{align}
 I(\omega)
   &=
   F\omega
   \prod_{j}
   \frac{P(\omega)}{\omega-\omega_j}
 \end{align}
 where $F$ is a positive constant and $P(\omega)$ an analytic function for complex frequencies (e.g., polynomial).
 We have seen earlier that, given a residue $\omega_j$, $\pm\omega_j^*$ are also residues of the bath spectral density.
 For purely imaginary residues $\omega_j$, the set of additional solutions reduces to $\tilde{\omega}_j^*=-\tilde{\omega}_j$.
 In this way, $P(\omega)$ becomes an even function of frequency as does $\prod_j (\omega-\omega_j)$.
 We note that $I(\omega)$ does not preserve causality under frequency integration (see also Equation \eqref{Eq:SpectralDen}).
 Causality, as we see in the following paragraph, is mathematically established by the convergence criterion of the Fourier transform in Equation~\eqref{Eq:NuGamma}.

 For illustration, let us consider the two bath spectral densities in Equations \eqref{Eq:ILambda} and \eqref{Eq:SpecDenOscillator}.
 Applying the residue theorem to the overdamped model in Equation \eqref{Eq:ILambda}, we obtain
 \begin{align}
 \label{Eq:NuLambdaModel}
 \nu_{\Lambda}(\tau)
   &=
   \frac{2}{\beta}
   \left(
   \gamma_{\Lambda}(\tau)
   +
   2
   \gamma_0\Lambda^2
   \sum_{n=1}^{\infty}
   \frac{\Lambda e^{-\Lambda|\tau|}-n\eta e^{- n\eta|\tau|}}{\Lambda^2-(n\eta )^2}
   \right)
 \end{align}
 which was, e.g., already reported in ref. \cite{einsiedler20}.
 For the resonance model $I_{\Omega}$ [Equation~\eqref{Eq:SpecDensApprox}], on the other hand, we find
 \begin{align}
 \label{Eq:NuOmegaModel}
 &\nu_{\Omega}(\tau)
   \sim
   \frac{2}{\beta}
   \gamma_{\Omega}(\tau)
   +
   \frac{4}{\beta}
   \rho(\mathbf{r}_a)
   \\\nonumber
   &\times
   \sum_{n=1}^{\infty}
   \left\{
   \frac{\Omega^2\cos[\Omega\tau]}{\Omega^2+[n\eta]^2}
   e^{-\frac{\kappa}{2}|\tau|}
   -
   \frac{n}{2}
   \frac{\eta\kappa \left(\Omega^2-[n\eta]^2\right)}{(\Omega^2+[n\eta]^2)^2}
   e^{-n\eta|\tau|}
   \right\},
 \end{align}
 where we assumed $\kappa/2\ll\Omega$ in the final expression.

 Together with the expression for the system's response function, an exact integration of the system's fluctuations $\langle\hat{Q}(t)\hat{Q}(t')\rangle_{\rm s}$ becomes possible for our choices of the spectral bath density.
 To this end, we first note that the polarizability $\alpha(\omega)$ features poles in the lower complex frequency plane only, in order to preserve causality. We can therefore evaluate the response function by means of the residue theorem and obtain
 \begin{align}
  \label{Eq:ResponseFunc}
 \chi(\tau)
 	&\sim
 	\int\frac{\mathrm{d}\omega}{2\pi}~
 	\frac{1}{\omega_0^2-\omega^2-2i\omega\gamma_{\theta}(\omega)}
 	e^{-i\omega \tau}
 	\\\nonumber
 	&
   =
 	\theta(\tau)
   \sum_{j=1}^N
   \prod_{k\neq j}
   \frac{P_1(\omega_j)}{\omega_j-\omega_k}
   e^{-i\omega_j\tau}
 	,
 \end{align}
 where $1\leq j\leq N$ is the number of roots $\omega_j$ solving
 \begin{align}
 [\omega_0^2-\omega_j^2]P_2(\omega_j)-2 i\omega_j P_1(\omega_j)=0
 \end{align}
 and we have written the causal dissipation kernel as the fraction of two complex polynomials $\gamma_{\theta}(\omega)=P_1(\omega)/P_2(\omega)$ (see, e.g., Equations \eqref{Eq:ILambda} and \eqref{Eq:SpecDensApprox} in combination with Equation \eqref{Eq:NuGamma}).
 $\chi(\tau)$ is a real function due to the crossing relation which translates into $\omega_j=-\omega_k^*$ for $k\neq j$ \cite{jackson99}.
 In other words, for purely complex resonances, as it is, e.g., the case in the model of Equation \eqref{Eq:ILambda}, the response function becomes exponentially damped. In the other case, i.e., the model in Equation \eqref{Eq:SpecDensApprox}, the system's response function generally behaves as the linear combination of damped oscillations with coupling constants (we follow the notation of ref. \cite{einsiedler20}).
 \begin{align}
 d_j
   &=
   \prod_{k\neq j}
   \frac{P_1(\omega_j)}{\omega_j-\omega_k}.
 \end{align}
 We iterate that the generalization to include branch cuts in the bath spectral density is possible but requires more care in the limiting procedures.

The properties of the field fluctuations $\nu$ and the system's response function $\chi$ are inherited by the fluctuations of the system degree of freedom, i.e.,
 $\langle\hat{Q}(t)\hat{Q}(t')\rangle_{\rm s}
   =
   \int_0^t\mathrm{d}x\int_0^{t'}\mathrm{d}y~
   \chi(x)
   \chi(y)
   \nu(y-x+t-t')
 $ and similarly for momentum and cross-correlations.
 For instance, in the case of the $\Lambda$-model, due to the relatively simple form of the response function $\chi$ and the noise correlations $\nu$ as a linear combination of exponentially damped functions, the corresponding time integrals can be solved analytically (see also ref. \cite{einsiedler20} for the case of $t=t'$).
 The only complexity arises from properly accounting for the time ordering between $t$ and $t'$.
 To this end, we note that only the modulus square of the time argument $\tau$ of the noise kernel (see Equations \eqref{Eq:NuLambdaModel} and \eqref{Eq:NuOmegaModel}) is relevant.
 For $t>t'$, we use that the argument of $\nu$ becomes negative (positive) for $y\leq(\geq)x-(t-t')$ and split the $y$-integral into $\int_0^{t'}\mathrm{d}y=\int_0^{x-(t-t')}\mathrm{d}y+\int_{x-(t-t')}^{t'}\mathrm{d}y$.
 For $t<t'$, we use that the noise kernel $\nu$ is symmetric in its argument and we reduce the problem to the previous case $t>t'$ by exchanging $t\leftrightarrow t'$.
 The subsequent integration is then straight-forward and we employ a computer algebra system to carry out the calculation.
The Mathematica code \cite{mathematica} and the calculated data is available upon reasonable request.
 Lastly, we numerically compute the sum over the coth's complex thermal resonances ($n\eta$ in Equation \eqref{Eq:NuLambdaModel}), which is converging rapidly (at least as $n^{-2}$).
 Taken together, the previous reasoning is the basis for the numerical evaluations performed for the present manuscript.

 \section{Current--Current Fluctuations in Steady-State Heat Transfer}\label{App:Heattrans}

 Given the system of coupled Langevin equations in Equation \eqref{Eq:LangevinCoupledOsc}, the autocorrelation reads at late times
 \begin{subequations}
 \label{Eq:SolCoupledOsc}
 \begin{align}
 \label{Eq:SolCoupledOsc1}
 \langle\hat{\bm{\chi}}^2(t)\rangle_{\rm s}
   &\sim
   \int\frac{\mathrm{d}\omega}{2\pi}~
   \underline{\mathbb{D}}(\omega)
   ,
   \\
 \underline{\mathbb{D}}(\omega)
   &=
   \underline{D}(\omega)
   \cdot
   \underline{G}_H(\omega)
   \cdot
   \underline{D}^{\dagger}(\omega),
   \\
 \underline{D}(\omega)
   &=
   \left[
   \underline{\Omega}-\omega^2-2i\gamma_0\omega
   \right]^{-1}
   .
 \end{align}
 \end{subequations}
 For further details, we also refer to refs. \cite{subasi12,pagel13,hsiang15}.
 Using Equation \eqref{Eq:SolCoupledOsc}, a straight-forward calculation yields for the first two moments in the steady state
 \begin{subequations}
 \begin{align}
 \frac{\langle\hat{J}\rangle_{\rm s}}{t}
   &\sim
   -
   \frac{\sigma}{2}
   \int\frac{\mathrm{d}\omega}{2\pi}~
   (i\omega)\left[\underline{\mathbb{D}}(\omega)\right]_{21},
   \\
   \label{Eq:JsquaredHeat}
 \frac{\langle(\Delta\hat{J})^2\rangle_{\rm s}}{t}
     &\sim
     -
     \frac{\sigma^2}{4}
     \int\frac{\mathrm{d}\omega}{2\pi}~
     \omega^2
     \left(
     \text{det}
     \left[\underline{D}(\omega)\cdot\underline{D}^{\dagger}(\omega)\right]
     \right.
     \\\nonumber
     &
     -
     \left[\underline{\mathbb{D}}(\omega)\right]_{22}
     \left[\underline{\mathbb{D}}(-\omega)\right]_{11}
     +
     \left[\underline{\mathbb{D}}(\omega)\right]_{21}
     \left[\underline{\mathbb{D}}(-\omega)\right]_{12}
     \Big),
 \end{align}
 \end{subequations}
 where the subscript denotes the indices of the matrix.
 In order to derive Equation \eqref{Eq:JsquaredHeat}, we reordered the operators using the quantum average of the commutator 
 \begin{align}
 \langle
 \left[
 \hat{\bm{\chi}}(s),
 \hat{\bm{\chi}}(\bar{s})
 \right]
 \rangle
   &=
   -\frac{\hbar}{4\pi}
   \int\frac{\mathrm{d}\omega}{2\pi}~
   \omega~
   \underline{D}(\omega)
   \cdot
   \underline{D}^{\dagger}(\omega)
   e^{-i\omega(s-\bar{s})}.
 \end{align}
 Furthermore, we used that elements of the form
 $
 \sum_{i,j=1,2}
 \langle\hat{\chi}_i(s)\hat{\chi}_j(\bar{s})\rangle_{\rm s}
 \langle[\hat{\chi}_j(s),\hat{\chi}_i(\bar{s})]\rangle
 $
 vanish under the integral at late times.
 Calculating the relevant matrix components, the previous result reduces to a Landauer-type formula for the average heat transfer
 \begin{align}
 \frac{\langle\hat{J}\rangle_{\rm s}}{t}
   &\sim
   -
   \int\frac{\mathrm{d}\omega}{2\pi}~
   (\hbar\omega)
   T(\omega)
   \\\nonumber
   &\times
   \left(
   \coth\left[\frac{\hbar\omega\beta_1}{2}\right]
   -
   \coth\left[\frac{\hbar\omega\beta_2}{2}\right]
   \right)
 \end{align}
with the transmission function defined below Equation \eqref{Eq:FlucHeatTransmission}.
 Similarly, we find Equation~\eqref{Eq:FlucHeatTransmission}.


\end{document}